\RequirePackage{fix-cm}
\documentclass[a4paper]{article}

\pdfoutput=1

\usepackage{jheppub}

\usepackage[utf8]{inputenc}
\usepackage[T1]{fontenc}
\usepackage[]{microtype}

\usepackage{amssymb,amsmath,amsfonts}
\usepackage{graphicx}
\usepackage{epsfig}
\usepackage[outdir=./]{epstopdf}
\usepackage{subfig}

\graphicspath{{figures/}}

\def\gosam{{\sc GoSam}}
\def\herwig{{\sc Herwig7}}
\def\matchbox{{\sc Matchbox}}
\def\hpp{{\sc Herwig++}}
\def\mcfm{{\sc MCFM}}

\newcommand{\be}{\begin{equation}}
\newcommand{\ee}{\end{equation}}

\def\ket#1{|{#1}\rangle}
\def\bra#1{\langle{#1}|}

\title{Anomalous coupling, top-mass and parton-shower effects in \boldmath${W^+W^-}$ production}

\author[a]{J.~Bellm,}
\author[b]{S.~Gieseke,}
\author[c]{N.~Greiner,}
\author[d]{G.~Heinrich,}
\author[a,e]{S.~Pl\"atzer,}
\author[b,f]{C.~Reuschle,}
\author[d]{J.F.~von~Soden-Fraunhofen}

\emailAdd{johannes.bellm@durham.ac.uk}
\emailAdd{stefan.gieseke@kit.edu}
\emailAdd{greiner@physik.uzh.ch}
\emailAdd{gudrun@mpp.mpg.de}
\emailAdd{simon.platzer@durham.ac.uk}
\emailAdd{creuschle@hep.fsu.edu}
\emailAdd{jfsoden@mpp.mpg.de}

\affiliation[a]{Institute for Particle Physics Phenomenology, Durham
  University, Durham DH1 3LE, UK}
\affiliation[b]{Institut f\"ur Theoretische Physik, Karlsruhe Institute of Technology, 76131 Karlsruhe, Germany}
\affiliation[c]{Physik-Institut, Universit\"at Z\"urich, Winterthurerstr. 190, 8057 Z\"urich, Switzerland }
\affiliation[d]{Max-Planck-Institut f\"ur Physik, F\"ohringer Ring 6, 80805 M\"unchen, Germany}
\affiliation[e]{Particle Physics Group, School of Physics and Astronomy, University of Manchester,
  Manchester M13 9PL, UK}
\affiliation[f]{HEP Theory Group, Department of Physics, Florida State University, Tallahassee, USA}


\abstract{We calculate the process $pp\to W^+W^-\to e^+ \nu_e
\mu^-\bar{\nu}_\mu$ at NLO QCD, 
including also effective field theory (EFT) operators mediating the $ggW^+W^-$ interaction, 
which first occur at dimension eight.
We further combine the NLO and EFT matrix elements produced by \gosam{} with the
\herwig{}/\matchbox{} framework,
which offers the possibility to study the impact of a parton shower.
We assess the effects of the anomalous couplings by comparing them to top-mass effects 
as well as uncertainties related to variations of the renormalisation, 
factorisation and hard shower scales.}

\preprint{          MPP-2016-14\\
\hspace*{0pt}\hfill KA-TP-06-2016\\
\hspace*{0pt}\hfill HERWIG-2016-02\\
\hspace*{0pt}\hfill IPPP/16/11\\
\hspace*{0pt}\hfill MAN/HEP/2016/04\\
\hspace*{0pt}\hfill ZU-TH-5/16\\
\hspace*{0pt}\hfill MCnet-16-05
}

\begin{document}

\maketitle

\section{Introduction}

Among the most important goals for the next phase of LHC data taking are precision tests 
of the electroweak symmetry breaking sector and the search for signs of new physics. 
In this respect, the final state of $W^+W^-$ plays a prominent role. 
For example, the continuum $pp\to W^+W^-\to l \bar{l}\nu \bar{\nu}$ is the dominant background 
in the measurement of $H\to W^+W^-\to l \bar{l}\nu \bar{\nu}$. 
The process $pp\to W^+W^-$ (+jets) also can be a major background for new-physics processes 
involving missing energy. 
Therefore it is very important to have good theoretical control
on the $pp \to W^+W^-$ cross section. 

Final states with two massive vector bosons recently  have attracted additional interest 
due to the fact that 
a slight excess at about 2\,TeV in the search for di-boson resonances 
has been reported by both ATLAS~\cite{Aad:2015owa,Aad:2015ufa} and
 CMS~\cite{Khachatryan:2014hpa,Khachatryan:2014gha}, 
most pronounced in the hadronic decay channel, which however does not
seem to persist in Run II.
Further, 
both the ATLAS and CMS measurements  for the $W^+W^-$ total
inclusive cross sections using the leptonic decay channels, at
7\,TeV~\cite{ATLAS:2012mec,Chatrchyan:2013yaa} 
and at 8\,TeV~\cite{ATLAS-CONF-2014-033,Chatrchyan:2013oev}, are about 10-20\%
higher than the NLO predictions obtained from 
{\tt MCFM}~\cite{Campbell:1999ah,Campbell:2011bn}, 
which include the $gg$-initiated sub-process~\cite{Campbell:2011cu}.
However, the latter discrepancy has been largely reduced by the NNLO predictions
which became available recently~\cite{Gehrmann:2014fva,Grazzini:2015wpa,Caola:2015rqy}. 
In addition, it has been noticed that resummation of large logarithms arising from the 
jet veto condition needs to be taken into account 
carefully~\cite{Moult:2014pja,Jaiswal:2014yba,Monni:2014zra,Jaiswal:2015vda}, 
and that the discrepancy for the  fiducial  cross section is only at the $1\sigma$ level, 
such that the way the extrapolation from the fiducial cross section is done
should be revisited~\cite{Monni:2014zra}.
Considering all these recent developments, 
the need for precise phenomenological studies, also at the level of differential distributions
and in view of possible BSM contributions, 
is evident.

Let us briefly review the history of higher-order calculations in the $W^+W^-$(+jets) channel:
The process $gg\to W^+W^-$ has been calculated in continuously improving approximations 
in the literature:
the calculation for on-shell $W$ bosons has been performed
in~\cite{Dicus:1987dj,Glover:1988fe}.
Leptonic decays of the $W$ bosons were included in~\cite{Binoth:2005ua} 
for massless fermion loops and extended to include the  masses of the
top and bottom quarks in~\cite{Binoth:2006mf}. 
Analytic results, including the mass of the top quark, were presented in~\cite{Campbell:2011bn,Campbell:2011cu}, 
together with a phenomenological study of interference effects with $H\to W^+W^-$. 
Focusing on a Higgs-boson mass of about  $125$\,GeV, an update of interference effects
has been performed in~\cite{Kauer:2012hd,Kauer:2013qba} and in~\cite{Bonvini:2013jha}, 
where the latter includes 
higher-order corrections to the interference in a soft-collinear approximation up to NNLO.
Very recently, the NNLO corrections to the process $pp\to W^+W^-$ were calculated in \cite{Gehrmann:2014fva}, removing the discrepancy to the data at 7\,TeV, 
and decreasing the excess at 8\,TeV to a level below $1\sigma$.
Electroweak corrections to the full 4-lepton final state, including also mass effects, 
have been calculated in \cite{Billoni:2013aba}. For a phenomenological study of 
electroweak and QCD effects see also \cite{Baglio:2013toa}. 
A study of combined electroweak \cite{Bierweiler:2013dja,Bierweiler:2012kw} and QCD corrections (assuming that they
factorise), 
including also matching to the angular-ordered parton shower, has been
performed in \cite{Gieseke:2014gka} and is also available in \herwig{}~\cite{Bahr:2008pv,Bellm:2015jjp}, 
the successor of \hpp~\cite{Hamilton:2010mb}.

The NLO QCD corrections to the process $q\bar{q}\to W^+W^-$ for
on-shell $W$ bosons have been calculated in~\cite{Ohnemus:1991kk,Frixione:1993yp}.
The helicity amplitudes for the process including decays 
have been calculated in~\cite{Dixon:1998py}, 
followed by phenomenological studies in~\cite{Dixon:1999di,Campbell:1999ah}. 
Matching with parton showers of these processes has been included in
{\sc MC@NLO} \cite{Frixione:2002ik}.
Weak-boson pair production with NLO QCD
corrections, matched to a parton shower with the {\sc Powheg} method
\cite{Alioli:2010xd,Frixione:2007vw}, has been directly implemented in
\hpp~\cite{Hamilton:2010mb}.

The process $pp\to H\to W^+W^-$ also has attracted recent interest in view of 
measuring the Higgs width using information from off-shell
production and decay, 
as proposed in \cite{Campbell:2013wga,Caola:2013yja,Kauer:2012hd}
and further investigated in \cite{Englert:2014aca,Buschmann:2014sia}.  
Such a measurement already has been performed based on the $ZZ$ final state~\cite{Khachatryan:2014iha}.

\medskip

Calculations of the process $pp\to W^+W^-+$\,jet, 
without including the $gg$ initial state, have been performed
 in~\cite{Dittmaier:2007th,Campbell:2007ev,Dittmaier:2009un}, 
and recently, including also NLO electroweak corrections,  in~\cite{Wei-Hua:2015gaa}.
The loop-induced process $gg\to W^+W^-+$\,jet has been studied in~\cite{Melia:2012zg}.
A very detailed NLO study of 4-lepton plus 0,1-jet final states, 
including NLO matching to a parton shower and merged samples, $H\to WW^*$ interference studies 
and squared quark-loop contributions, has been presented in \cite{Cascioli:2013gfa}.
 
\medskip

In this paper, we calculate the process $pp\;(\to W^+W^-)\to e^+ \nu_e
\mu^-\bar{\nu}_\mu$ at NLO QCD, combining the hard matrix elements produced
by \gosam~\cite{Cullen:2011ac,Cullen:2014yla} with the
\herwig{}/\matchbox~\cite{Bahr:2008pv,Platzer:2011bc,Bellm:2015jjp} framework,
which offers the possibility to study the impact of a parton shower.  In
addition, we particularly focus on the loop-induced process $gg\;(\to W^+W^-)\to e^+ \nu_e
\mu^-\bar{\nu}_\mu$, where we investigate how new-physics effects which
modify the effective $gg\,W^+W^-$ coupling could affect various distributions. To
this aim we include the most general effective field theory (EFT) operators mediating the $gg\,W^+W^-$
interaction, which first occur at dimension eight, in our automated setup.
This allows us to assess the impact of these operators in various effective
coupling scenarios.

\vspace{15pt}
\section{Details of the calculation}

\subsection{The loop-induced contribution $gg\to e^+ \nu_e \mu^-\bar{\nu}_\mu$}

\begin{figure}[h!]
\begin{center}
\includegraphics[width=0.75\textwidth]{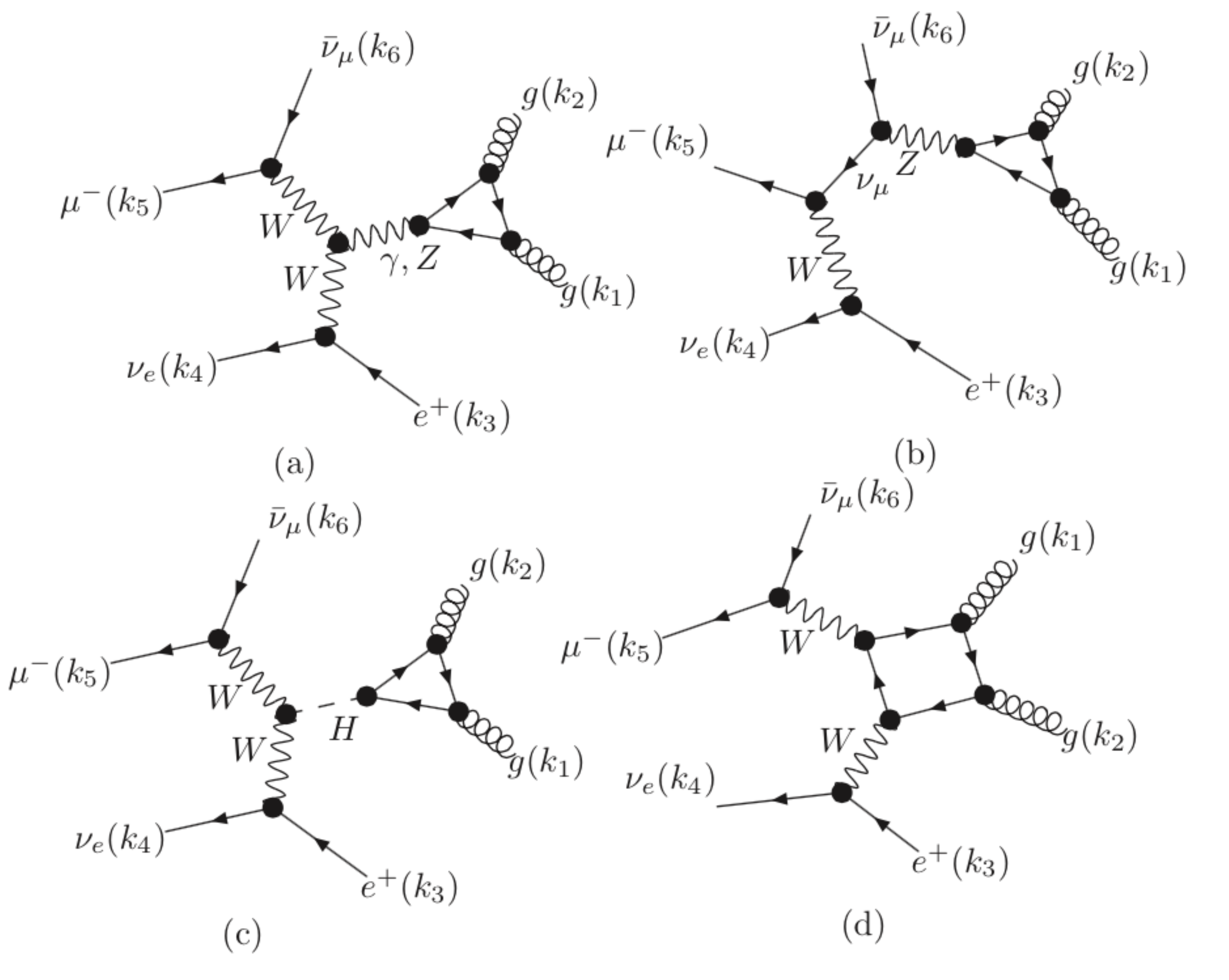}
\end{center}
\caption{Examples of diagrams contributing to the process $gg\to e^+ \nu_e \mu^-\bar{\nu}_\mu$.}
\label{fig:diagrams}
\end{figure}

The diagrams contributing to the process $gg\to e^+ \nu_e
\mu^-\bar{\nu}_\mu$ (see Fig.~\ref{fig:diagrams})
comprise of diagrams involving two $W$ propagators (``doubly resonant'') as well as diagrams
involving only one $W$ propagator (``singly resonant'', see Fig.~\ref{fig:diagrams}b).
Note that the latter are
important to maintain gauge invariance.
Non-resonant diagrams, {\it i.e.} diagrams containing no $W$ propagator, do
not contribute to the $e^+ \nu_e \mu^-\bar{\nu}_\mu$ final state, but
they would contribute to the 
$e^+ \nu_e e^-\bar{\nu}_e$ final state, which we are not considering here.

We include massive top- and bottom-quark loops, all other quarks (and leptons) 
in the hard process are assumed to be massless.
The photon-exchange graphs vanish due to Furry's theorem.
The $Z$-exchange diagrams are proportional to
$(m_u^2-m_d^2)(p_3^2-p_4^2)$, when summed over up- and down-type contributions.
Therefore these diagrams also vanish for massless quarks.
For arbitrary invariant masses of the charged\,lepton\,-\,neutrino pairs, {\it i.e.} $p_3^2\neq p_4^2$, 
and the third generation, 
where we assume $m_t\not=0$ and $m_b\not=0$ in the loops, 
we find that the contributions from doubly-resonant (Fig.~\ref{fig:diagrams}a) 
and singly-resonant (Fig.~\ref{fig:diagrams}b) diagrams with internal $Z$-propagator cancel each other. 
The only triangle graphs that contribute are thus the Higgs-exchange
diagrams (Fig.~\ref{fig:diagrams}c) 
where the amplitude contains  the $Q\bar{Q}H$ Yukawa couplings.
The box diagrams do not involve these couplings
and therefore form a gauge-invariant subset. 

\vspace{15pt}
\subsection{Operators parametrizing the $gg\,W^+W^-$ coupling}
\label{sec:eftoperators}

The first operators that mediate a four-boson interaction between the two gluons and the two $W$ bosons
occur at dimension eight.

The gluonic field-strength tensor is defined as
\begin{equation}
 G^a_{\mu\nu}=\partial_{\mu}G^a_{\nu} -\partial_{\nu}G^a_{\mu}-g_s f^{abc}G^b_{\mu}G^c_{\nu}\;,
\end{equation}
the field-strength tensor of the $W$ is defined as
\begin{equation}
 W^{I}_{\mu\nu}= \partial_{\mu}W^I_{\nu} -\partial_{\nu}W^I_{\mu}-g \epsilon^{IJK}W^J_{\mu}W^K_{\nu},\;\;I,J,K\in \{1,\ldots,3\}\;.
\end{equation}

We can write the SU(2) fields $W^{I}$ in terms of the physical fields:
\begin{flalign}
 W^1_{\mu}&=\frac{1}{\sqrt{2}}\left(W^{+}_{\mu}+W^{-}_{\mu}\right)\nonumber \\
 W^2_{\mu}&=\frac{i}{\sqrt{2}}\left(W^{+}_{\mu}-W^{-}_{\mu}\right)\nonumber \\
 W^3_{\mu}&=Z_{\mu}\cos\theta_w +A_{\mu}\sin\theta_w\;.
\end{flalign}

A CP-odd operator can be introduced via the dual field-strength tensors which are defined by
\begin{equation}
\tilde{G}^{a,\mu\nu} =\frac{1}{2}\epsilon^{\mu\nu\rho\sigma}G^a_{\rho\sigma}\;,\quad
\tilde{W}^{I,\mu\nu}
=\frac{1}{2}\epsilon^{\mu\nu\rho\sigma}W^I_{\rho\sigma}\;\ .
\end{equation}

Based on these field-strength tensors we can build the dimension-eight operators contributing to the Lagrangian:
\begin{flalign}
 {\cal O}_1&=
 \frac{c_1}{\Lambda^4}G^a_{\mu\nu}G^{a,\mu\nu}W^I_{\rho\sigma}W^{I,\rho\sigma}\;=\frac{c_1}{\Lambda^4}\,\tilde{{\cal
   O}}_1\nonumber \\
 {\cal O}_2&= \frac{c_2}{\Lambda^4}\tilde{G}^a_{\mu\nu}G^{a,\mu\nu}W^I_{\rho\sigma}W^{I,\rho\sigma}\;= \frac{c_2}{\Lambda^4}\,\tilde{{\cal
   O}}_2\nonumber \\
 {\cal O}_3&= \frac{c_3}{\Lambda^4}G^a_{\mu\nu}G^{a,\mu\nu}\tilde{W}^I_{\rho\sigma}W^{I,\rho\sigma}\;=\frac{c_3}{\Lambda^4}\,\tilde{{\cal
   O}}_3\;\ ,
\end{flalign}
where $\Lambda$ denotes the scale below which the EFT description is valid.

Combining the SM Lagrangian with the one including the effective couplings, we have
\be
{\cal L}_{eff}={\cal L}_{SM}+\frac{1}{\Lambda^4}\sum_{i} c_i\,\tilde{{\cal
    O}_i}\;.
\ee

Combining higher-order QCD corrections from ${\cal L}_{SM}$ with the part
containing the higher-dimensional operators means that we are
performing a simultaneous expansion in $\alpha_s/2\pi$ and in
$c_i/\Lambda^4$.
This requires a careful assessment of the relative importance of the
various terms in such an expansion and of the range of validity of the
effective theory. We will come back to this issue in Section
\ref{sec:pheno}. 

\begin{figure}[htb]
\centering
\subfloat[\label{fig:MEsq}] {\includegraphics[width=0.23\textwidth]{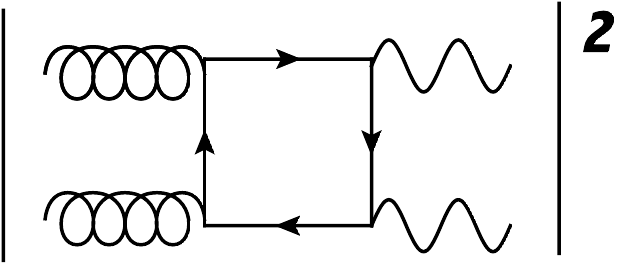}}\hspace*{7.5mm}
\subfloat[\label{fig:interference}]{\includegraphics[width=0.4\textwidth]{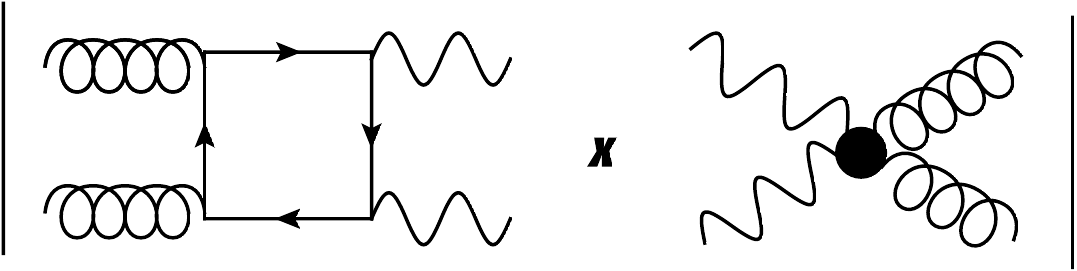}}\hspace*{12.5mm}
\subfloat[\label{fig:dim8sq}]{\includegraphics[width=0.18\textwidth]{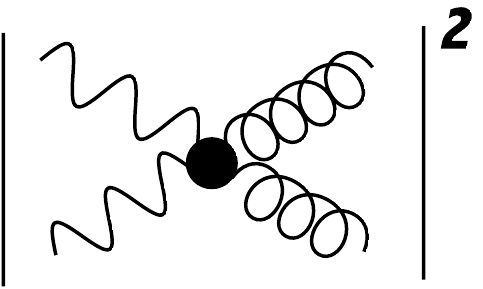}}
\caption{The three types of contributions to the squared matrix element.}
\label{fig:MEcontribs}
\end{figure}

As in the SM the $gg\,W^+W^-$ coupling is loop induced, we will calculate the
following contributions to the $gg\;(\to W^+W^-)\to e^+ \nu_e \mu^-\bar{\nu}_\mu$ cross section, depicted
schematically in Fig.~\ref{fig:MEcontribs}:
\be
\sigma_{ggWW}\sim |{\cal M}_{\mathrm{SM}}^{\text{1-loop}}|^2+2\,\mathrm{Re} \left( {\cal
    M}_{\mathrm{SM}}^{\text{1-loop}}\,{\cal M}_{\text{dim-8}}^*\right)  +|{\cal M}_{\text{dim-8}}|^2\;.
\label{eq:MEsq}
\ee

Note that the last term above is suppressed by $1/\Lambda^8$.

In the following we list the Feynman rules corresponding to the
dimension-eight operators. All momenta are considered to be incoming:

\vspace{5pt}
\begin{minipage}{4.5cm}
\center
\vspace{5pt}
\hspace{15pt}
\includegraphics[scale=0.33]{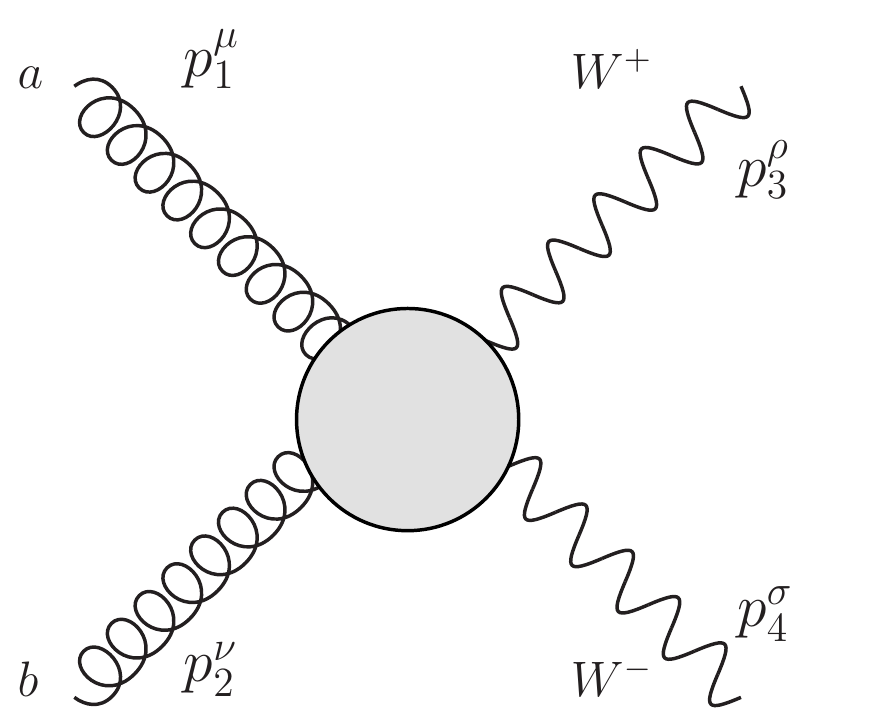}
\end{minipage}
\begin{minipage}{9.9cm}
\vspace{-5pt}
\begin{flalign}
 {\cal O}_1:&\quad 16i\frac{c_1}{\Lambda^4}\delta^{a,b}\left(p_1^{\nu}p_2^{\mu} -g^{\mu\nu}p_1\cdot p_2\right)
\left(p_3^{\sigma}p_4^{\rho} -g^{\rho\sigma}p_3\cdot p_4\right) \nonumber \\
 {\cal O}_2:&\quad 16i\frac{c_2}{\Lambda^4}\delta^{a,b}\epsilon^{\mu \nu p_1 p_2}\left(p_4^{\rho} p_3^{\sigma} -g^{\rho \sigma}p_3 \cdot p_4\right) \nonumber \\
{\cal O}_3:&\quad 16i\frac{c_3}{\Lambda^4}\delta^{a,b}\epsilon^{\rho \sigma p_3 p_4}\left(p_1^{\nu} p_2^{\mu} -g^{\mu \nu}p_1 \cdot p_2\right)
\end{flalign}
\end{minipage}

\vspace{15pt}
\subsection{Loop-induced processes in \gosam}

The virtual amplitudes for the process  $pp\;(\to W^+W^-)\to e^+ \nu_e \mu^-\bar{\nu}_\mu$,
as well as the spin- and colour-correlated tree amplitudes,
have been generated with the program \gosam~\cite{Cullen:2011ac,Cullen:2014yla},
which is an automated package to generate one-loop amplitudes. It is 
based on a Feynman-diagrammatic approach using {\sc{Qgraf}}~\cite{Nogueira:1991ex} and
{\sc{Form}}~\cite{Kuipers:2012rf,Kuipers:2013pba} for the diagram generation, and
{\sc{Spinney}}~\cite{Cullen:2010jv} and
{\sc{Form}} to produce optimised {\sc Fortran90} code. For the reduction of the
one-loop amplitudes the user can choose between three different reduction libraries 
(or a combination thereof):
{\sc{Ninja}}~\cite{Mastrolia:2012bu,vanDeurzen:2013saa,Peraro:2014cba},
{\sc{Golem95}}~\cite{Binoth:2008uq,Cullen:2011kv,Guillet:2013msa} or {\sc{Samurai}}~\cite{Mastrolia:2010nb,vanDeurzen:2013pja}.
The default setup uses {\sc{Ninja}} in combination with {\sc{Golem95}} as a rescue system for 
numerically problematic phase-space points.
The scalar basis integrals
have been evaluated using {\sc{Golem95}} and {\sc{OneLOop}}~\cite{vanHameren:2010cp}.  
We use the complex mass scheme~\cite{Denner:2006ic} throughout our calculation.

\medskip

The calculation of loop-induced processes is straightforward with \gosam, 
as it is based on a Feynman-diagrammatic approach. 
However, we have improved the rescue system for loop-induced processes: 
in the standard case where a tree-level amplitude exists, the accuracy at a given phase-space point 
is assessed by comparing the 
value for the coefficient of the single infrared (IR) pole with the exact value (which is obtained from the universal IR 
behaviour of the subtraction terms). As this coefficient must be zero in the loop-induced case, 
we have implemented an accuracy check which tests the precision of this zero.
In more detail, we have added the following options to the \gosam{} input card:
\begin{itemize}
\item \texttt{PSP\_chk\_li1}: allows to set the desired precision of the pole part (which should be zero) in
comparison to the finite part. If the pole part is at least
\texttt{PSP\_chk\_li1} orders smaller than the finite part, the point is
accepted.
\item \texttt{PSP\_chk\_li2}:  for loop-induced
processes, this option is used instead of \texttt{PSP\_chk\_th2}.
It is the threshold to declare a phase-space point  as ``bad'', based on the precision
of the pole in comparison to the finite part. Points with
precision less than this threshold are directly reprocessed with
the rescue system (if switched on), or declared as unstable.
According to the verbosity level set, such points are written to
a file and not used when the code is interfaced to an external
Monte Carlo program in accordance with the updated BLHA standard~\cite{Alioli:2013nda}.
\item Similarly, \texttt{PSP\_chk\_li3} is used instead of \texttt{PSP\_chk\_th3} as threshold for
      the rotation test in the loop-induced case.
\item \texttt{PSP\_chk\_li4} sets the minimum pole precision for the points
which already have been reprocessed by the rescue system. If a rescued point 
gives a pole coefficient which is at least
\texttt{PSP\_chk\_li4} orders smaller than the finite part, the point is
accepted.
\end{itemize}

\noindent For the Standard-Model case, we have validated our results by comparing to \mcfm~\cite{Campbell:2011cu}.

\vspace{15pt}
\subsection{Extended BSM support in \gosam{}}

We have implemented various new features in \gosam{} which facilitate the calculation of 
corrections beyond the Standard Model, as well as the interference between SM loop corrections and 
BSM effects described within an EFT framework. 
Among the new features are:
\begin{itemize}
\item the import of BSM model files in UFO format~\cite{Degrande:2011ua} in combination with 
the updated BLHA standard~\cite{Alioli:2013nda} for the definition of new couplings has been implemented,
\item BSM-SM interference terms can be calculated by specifying the orders in the corresponding couplings,
\item sub-process specific settings in the \gosam{} input card are possible,
\item the renormalisation term for the Wilson coefficient in the effective ggH coupling has been added,
\item the coefficients of the effective couplings multiplying the higher-dimensional operators can be 
modified by the user in the \gosam{} input card.
\end{itemize}

For the process considered here we use a model file in UFO format~\cite{Degrande:2011ua}
which we create by extending the SM Lagrangian of {\sc FeynRules}~\cite{Alloul:2013bka}
with the EFT operators outlined in Sec.~\ref{sec:eftoperators}. The SM parameters,
which are provided per default with the UFO model file, are then overwritten in accordance to the updated BLHA
standard~\cite{Alioli:2013nda} by the ones listed in Sec.~\ref{sec:pheno}.
In order to be able to compute the pure
SM NLO QCD contributions as well as the additional interference between the SM one-loop contribution and the effective
coupling with only one model file, we restrict the one-loop contributions
to allow only for SM couplings, with the help of \gosam's diagram filter facilities, while the tree-level contributions
are allowed to include the effective coupling.


\vspace{15pt}
\subsection{Interface to \herwig{} and computational setup}

\herwig{} features the full simulation of particle collision events up to the
particle level, {\it i.e.} perturbative as well as non-perturbative physics. The
perturbative part provides the simulation of
hard processes at NLO QCD (including several built-in LO and NLO matrix
elements, LH event file input as well as the fully automated assembly of NLO
QCD calculations for almost all Standard-Model processes, utilizing various
interfaces to several external matrix-element providers),
shower Monte Carlo algorithms,
as well as the corresponding LO and NLO matching procedures (dedicated matrix
element corrected shower plug-ins and built-in matched cross sections,
as well as a fully automated matching machinery).

The capabilities of \herwig{} for integration, unweighting and sub-process parallelization,
as well as the steering at the level of input files, are significantly improved.
By virtue of the \matchbox{} framework 
new integrator modules were introduced,
which provide for an efficient, automated multi-channel phase-space sampling:
the one which we employ, for the process considered here, is based on the standard sampling algorithm
contained in the ExSample library~\cite{Platzer:2011dr}.

Based on the \matchbox{} framework, \herwig{} facilitates the
automated setup of all sub-processes and ingredients necessary for a full NLO QCD
calculation in the subtraction formalism:
an implementation of the dipole subtraction method based on the approach by
Catani and Seymour~\cite{Catani:1996vz},
interfaces to various external matrix-element providers,
or plug-ins to various in-house calculations for the hard sub-processes
or to built-in colour and helicity sub-amplitudes.

In the case where the necessary one-loop and tree-level parts
are obtained from external matrix-element providers there exist several
possibilities:
either at the level of colour-ordered sub-amplitudes or
at the level of squared matrix elements 
through the updated BLHA standard interface~\cite{Alioli:2013nda}.
A more detailed description  of the interface
between {\sc Herwig++} and \gosam{}
has been given in \cite{Andersen:2014efa}, for the example of $Z+$jet production.
The interface between \herwig{}/\matchbox~\cite{Platzer:2011bc,Bellm:2015jjp}
and \gosam{\sc-2.0}~\cite{Cullen:2014yla} is fully automated for one-loop QCD
corrections,
and extensions thereof to handle
loop-induced processes and  additional contributions from EFT operators
are employed for the process considered here.

Fully automated matching algorithms are available, inspired by
MC@NLO~\cite{Frixione:2002ik} and Powheg~\cite{Nason:2004rx} (referred
to as subtractive and multiplicative matching respectively), for the systematic
and consistent combination of NLO QCD calculations with both shower variants in
\herwig{} (facilitating two coherent shower algorithms - an
angular-ordered parton shower~\cite{Gieseke:2003rz} as well as a dipole
shower~\cite{Platzer:2009jq}, including the simulation of decays with full spin
correlations).
For the process studied here, we eventually combine our fixed-order NLO
result (supplemented with loop-induced and EFT contributions)
with the \herwig{} angular-ordered parton shower \cite{Gieseke:2003rz}
through the subtractive ({\it i.e.} MC@NLO-like) matching algorithm based on
\cite{Frixione:2002ik,Platzer:2011bc}.

\vspace{15pt}
\section{Phenomenological studies}
\label{sec:pheno}

For our phenomenological studies of the process $pp \to e^+ \nu_e \mu^-\bar{\nu}_\mu + X$ 
we choose a center-of-mass energy of 13\,TeV.
We also study the behaviour of the BSM effects  when going from  8\,TeV to 13\,TeV.

We use the MMHT2014nlo68c\_nf4~\cite{Harland-Lang:2014zoa} parton distribution functions (PDFs),
with 4 massless 
quark flavours\footnote{
For the b quarks circulating in the loops we use $m_b=4.2$\,GeV. 
We have found that the effect of finite b-quark masses in the loops is below 0.1\%.
} in the initial state,
and we set $\alpha_s(M_Z=91.1876\,\text{GeV})=0.12$ in accordance with the PDF set we use.

Our default scale choice for the 13\,TeV results is a dynamic scale, 
$\mu_r=\mu_F=m_{WW}=\sqrt{(p_{e^+}+p_{\mu^-}+p_{\nu_e}+p_{\bar{\nu}_\mu})^2}$.
For comparisons we also use a fixed scale, $\mu_r=\mu_F=M_W$.

The mass of the top quark has been set to $M_t=174.2$\,GeV,
the Higgs mass to $M_H=125.7$\,GeV.
We further use a non-zero top width of $\Gamma_t=1.4$\,GeV,
and a Higgs width of $\Gamma_H=4.11$\,MeV.

For the electroweak input parameters we follow the SLHA~\cite{Skands:2003cj}
scheme\footnote{
This scheme is frequently used per default by many {\sc FeynRules}/UFO models.
}
for a set of SM low-scale input parameters to fix
the electroweak sector: the electroweak input parameters we choose are
$M_Z=91.1876$\,GeV,
$\alpha=\alpha(M_Z)=1/128.91$ and
$G_F=1.16637\cdot 10^{-5}$\,GeV${}^{-2}$,
from which $M_W$ and $\sin^2(\theta_w)$ are subsequently derived.
Furthermore we choose
$\Gamma_Z=2.4952$\,GeV and
$\Gamma_W=2.085$\,GeV.

The events are analysed using an in-house analysis for {\sc Rivet-2.4} \cite{Buckley:2010ar} interfaced by \herwig{}.
The $W$ bosons are directly reconstructed from their leptonic decay products
(not via the built-in \texttt{WFinder} function of {\sc Rivet}).

Our cuts on the analysis level are as follows.
To mimic $W$-identification cuts, we employ a cut on the invariant mass of each
same flavour lepton-neutrino pair of $60\,{\mathrm{GeV}}\leq m_{l\nu_l}\leq 100\,{\mathrm{GeV}}$. 
We further require the net transverse momentum of the lepton-neutrino pair to be larger than 10\,GeV, 
and a minimum $p_T$ of 25\,GeV for each identified lepton and 
for the missing transverse energy of the event.
The identified leptons are required to be in the rapidity range $-3\leq y_l\leq 3$.

For numerical stability we also employ cuts at the generator level, which are of course
less restrictive than the cuts employed at the analysis level.

\medskip

In the following three subsections we will first concentrate on the fixed-order results,
and then discuss the impact of a parton shower in Sec.~\ref{sec:ps}.

\vspace{15pt}
\subsection{Gluon-induced contributions and effects of higher-dimensional operators}
\label{sec:gg_parton}

We start the discussion of the results with the $gg$-initiated processes, {\it i.e.} contributions which are either
loop induced in the SM, or require dimension-eight operators to
contribute at the tree level. 
In the following
we will distinguish three different contributions, see Eq.~(\ref{eq:MEsq}).
One is the pure Standard-Model loop-induced contribution, where
the matrix element is the square of the $gg$-initiated one-loop amplitude. In the plots this contribution is denoted as {\tt{gg\_SM}}.
The second contribution is the interference term between the SM one-loop amplitude and the EFT tree-level
amplitude, which means it is a linear term in the higher-dimensional operators.
 This contribution is labeled as {\tt{gg\_Interf}}. 
Finally the third contribution stems from the square of the
EFT tree-level amplitude and is therefore quadratic in the
higher-dimensional operators. This contribution is labeled as
{\tt{gg\_Eff2}}.
This distinction allows us to separate the term linear in the higher-dimensional operators from the quadratic one
and study their behaviour independently.

Consequently \texttt{gg\_SM+Interf} and \texttt{gg\_Eff2+Interf} denote the combination of
\texttt{gg\_SM} or \texttt{gg\_Eff2} with \texttt{gg\_Interf} respectively,
while \texttt{gg\_All} finally denotes the combination of all contributions to the $gg$ initial state
({\it cf.} Fig.~\ref{fig:MEcontribs} or Eq.~(\ref{eq:MEsq})).

For the numerical results we have set 
\begin{equation}
 \frac{c_1}{\Lambda^4}=\frac{c_2}{\Lambda^4}=\frac{c_3}{\Lambda^4}= 0.1\,\text{TeV}^{-4}\;,
\end{equation}
unless stated otherwise.
\begin{figure}[htb]
\centering
 \subfloat[\label{fig:mww_gg_sm_bsm}]{ \includegraphics[width=0.45\textwidth]{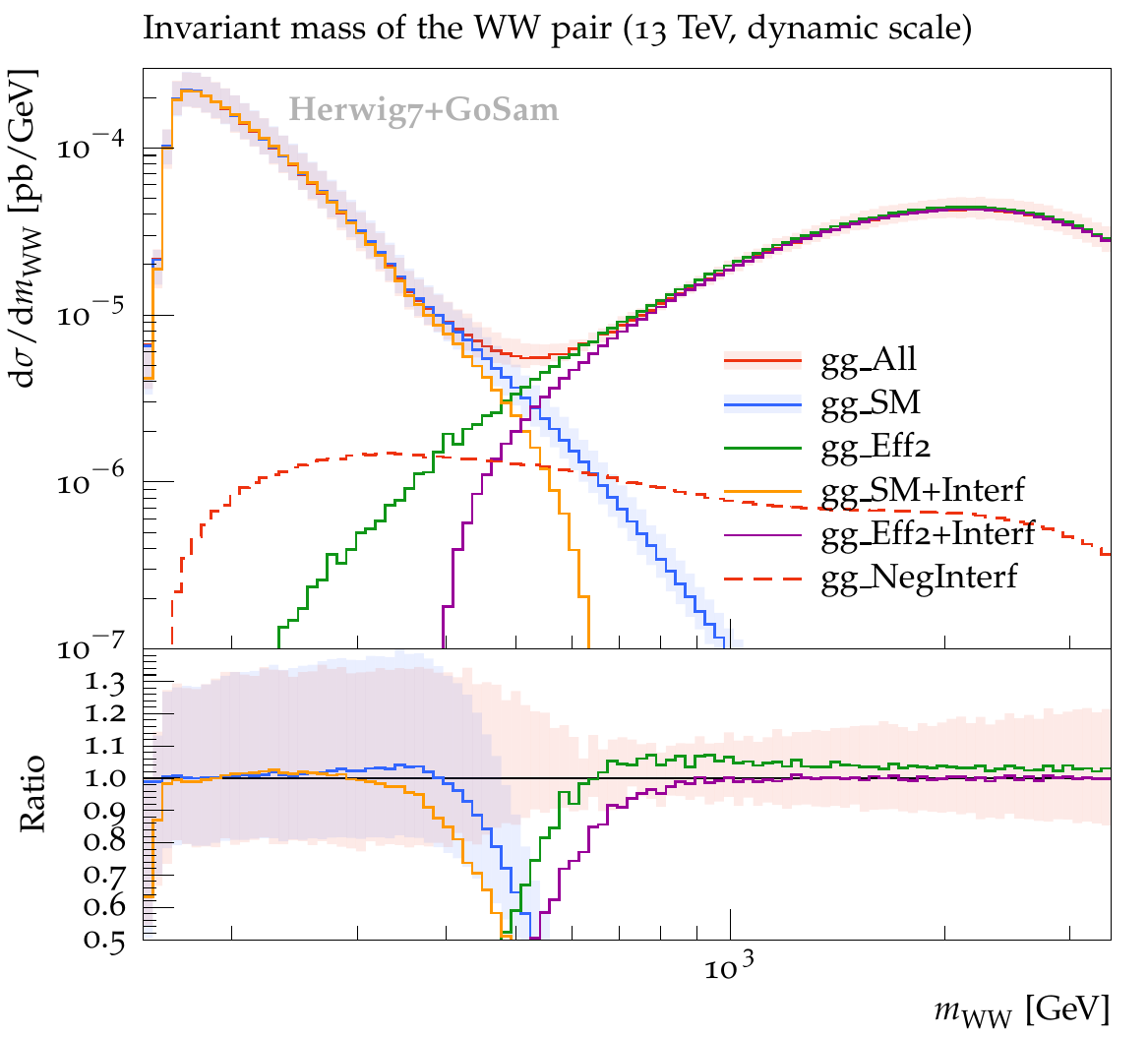}} \hfill
 \subfloat[\label{fig:deltaRww_gg_sm_bsm}]{ \includegraphics[width=0.45\textwidth]{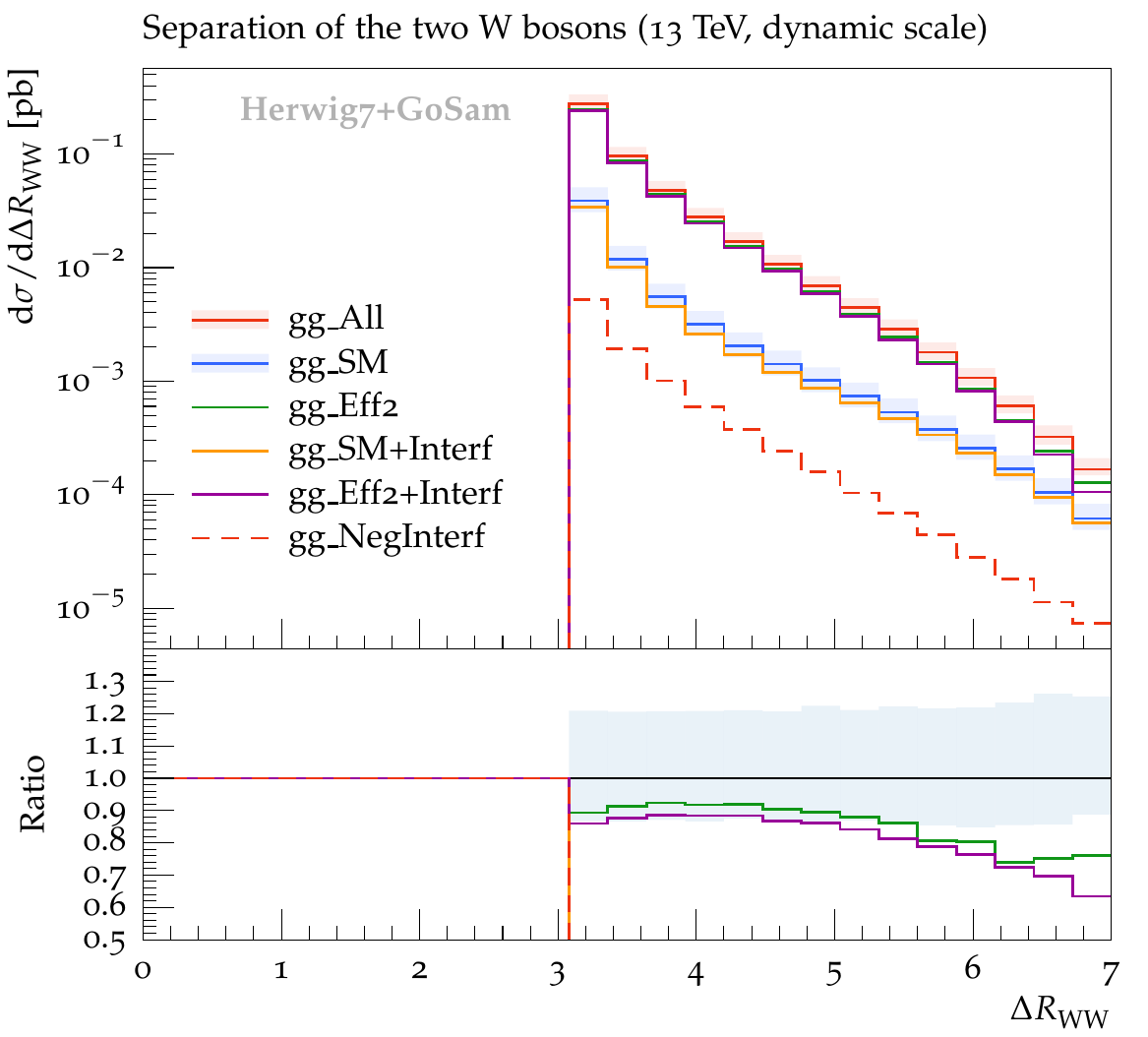}}
\caption{(a) $W$-boson pair invariant-mass distribution and (b) $\Delta
  R_{WW}$ distribution for the SM/BSM  $gg$-initiated
  contributions, at $\sqrt{s}=13$\,TeV. The shaded bands show the
  scale-variation uncertainties.
Ratio plots are with respect to \texttt{gg\_All}.
} 
\end{figure}
\begin{figure}[htb]
\centering
 \subfloat[\label{fig:mww_gg_sm_bsm_8Tev}]{ \includegraphics[width=0.45\textwidth]{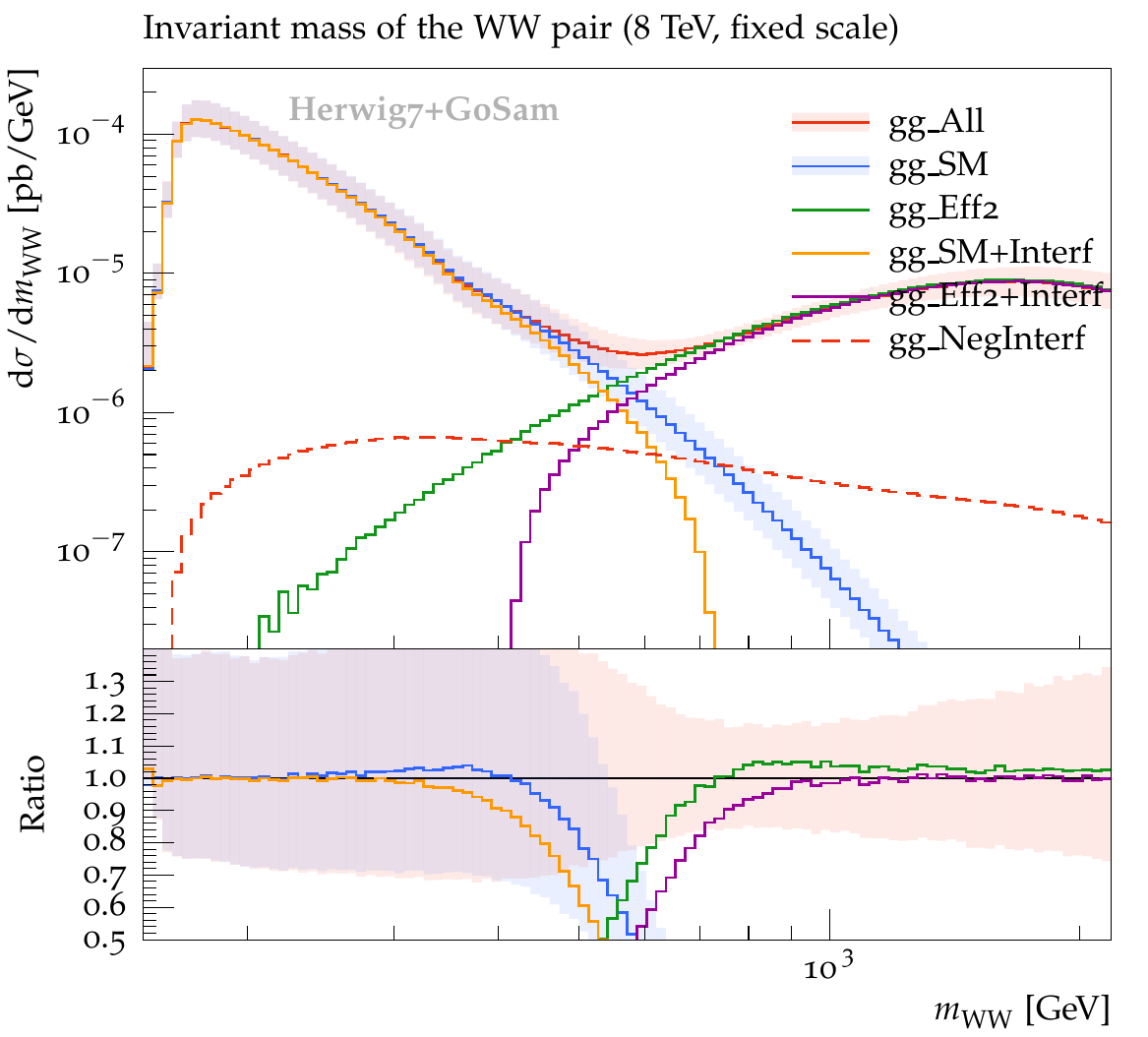}} \hfill
 \subfloat[\label{fig:deltaRww_gg_sm_bsm_8Tev}]{ \includegraphics[width=0.45\textwidth]{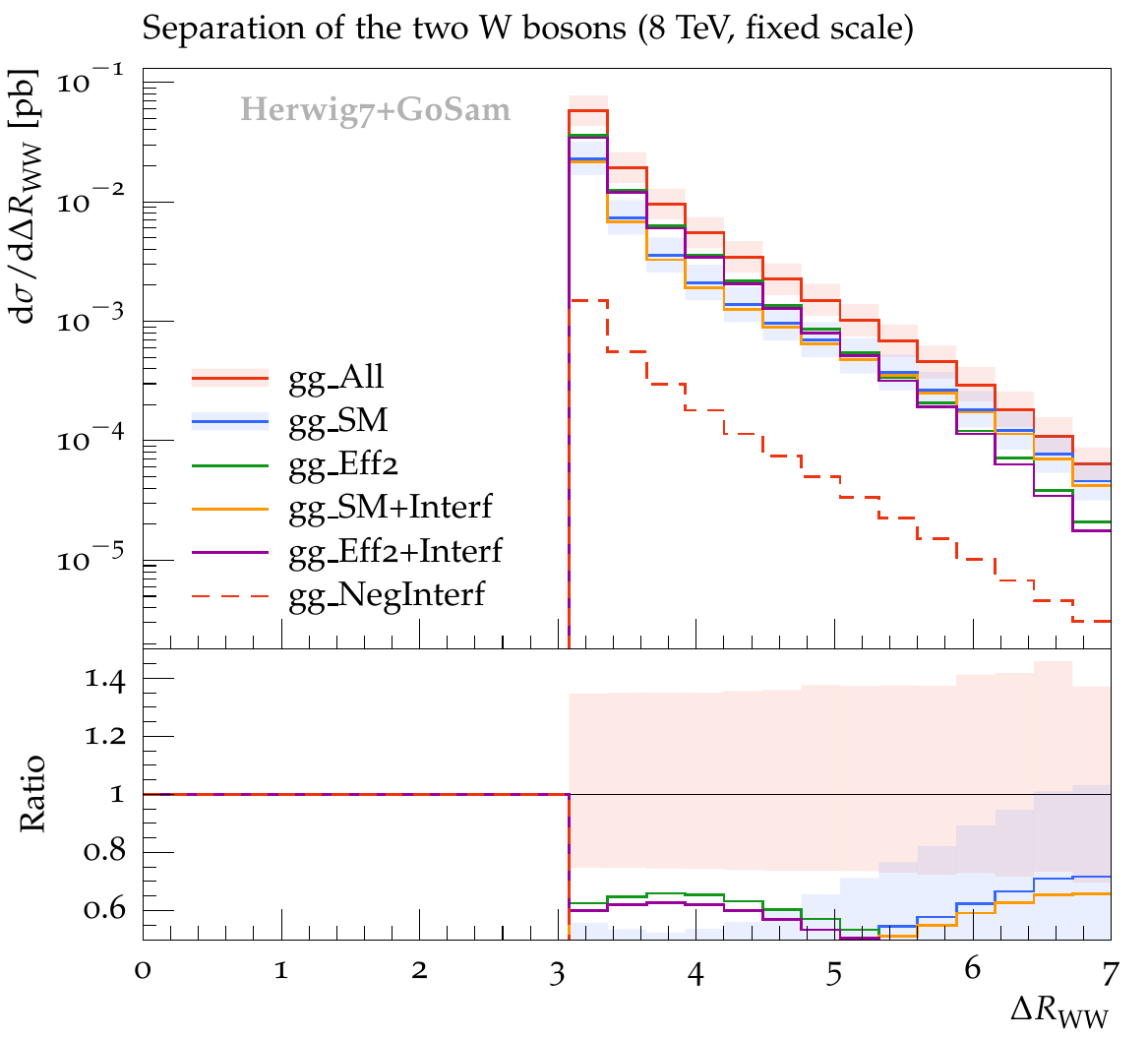}}
\caption{(a) $W$-boson pair invariant-mass distribution and (b) $\Delta
  R_{WW}$ distribution for the SM/BSM  $gg$-initiated
  contributions, at $\sqrt{s}=8$\,TeV.
The shaded bands show the scale-variation uncertainties. Ratio plots are with respect to \texttt{gg\_All}.
} 
\end{figure}

Let us first focus on the invariant-mass distribution of the $W$-boson pair, shown in Fig.~\ref{fig:mww_gg_sm_bsm}.
The invariant mass is defined via the momenta of the decay products:
\begin{equation}
 m_{WW} = \sqrt{(p_{e^{+}}+p_{\nu_e}+p_{\mu^{-}}+p_{\bar{\nu}_{\mu}})^2}\;.
\end{equation}

The most striking feature is the fact that the interference term of the SM loop-induced $gg\to W^+W^-$
 amplitude with the amplitude induced by the dimension-eight operators is negative. As this cannot be 
displayed in a logarithmic plot, we display it with the sign switched, denoted by {\tt{gg\_NegInterf}}.
As expected, the term linear in the dimension-eight operators dominates over the pure EFT contribution
(\texttt{gg\_Eff2}) which is 
quadratic in these operators. This is an obvious behaviour as the
quadratic terms receive an additional suppression of
a factor $c_i/\Lambda^4$. However its contribution increases with the center-of-mass energy (here $\sqrt{\hat{s}}=m_{WW}$)
and will eventually dominate over the linear term. The point where this happens depends on the setup 
and in particular on the chosen value for the anomalous coupling constants. In our example this happens 
at about $500$\,GeV (where the yellow and purple curves cross). As the linear term is negative, this is related to the point where the sum of the two higher-dimensional
contributions (\texttt{gg\_Eff2+Interf}) becomes positive, which happens a bit earlier at about 400\,GeV (where the green and the dashed red curves cross).
While the SM contribution drops rapidly as $m_{WW}$ is increasing, the dimension-eight contributions 
increase and start to dominate at around $m_{WW}\sim 500$\,GeV.
This is the expected behaviour as the contribution from a dimension-eight operator can increase maximally 
with $s^2/\Lambda^4$.

The scale-variation uncertainties (shown as shaded bands in the plots) are
relatively large, due to the fact that
the results for the $gg$-initiated subprocess are leading-order
predictions.

The fact that the term linear in the dimension-eight operators is negative leaves us with a phenomenologically
interesting constellation.
In the region where the linear term is dominant we get a decrease in the cross section compared to the SM
prediction, and with increasing invariant mass we observe a (partial) cancellation between the linear and
the quadratic term. At the point where linear and quadratic term are of the same magnitude, we recover exactly
the Standard-Model contribution. This means that
putting experimental constraints on these
types of couplings will be more difficult, 
and signs of new physics would be masked:
in the low energy region the effects of the 
dimension-eight operators are anyway suppressed by a factor of $1/\Lambda^4$, and for larger energies we find
(partial) cancellation between the linear and the quadratic term. However, we also emphasize that the sign
of the dimension-eight operators is not necessarily fixed to be positive. A negative value would lead to a
constructive interference between the linear and the quadratic term rather than to a cancellation.
The bounds on negative values will therefore be much more restrictive than on positive values.

The region where the quadratic term becomes equally important and eventually dominates over the linear
term has to be interpreted with care. The two terms being equally important means that the suppression of
the quadratic terms by the additional factor $c_i/\Lambda^4$ is compensated by a factor of $s^2$.
In other words $s\sim \Lambda^2$, which means that we are probing the scale of New Physics
and which is the point where the EFT approach becomes invalid, as it is based
on the assumption that it is a low-energy effective theory and that the scale of New Physics is much
higher than the scale we are probing.
It is this assumption that allows us to be confident 
that operators of lower dimensions are more important compared to higher-dimensional operators.
If higher-dimensional operators were  not sufficiently suppressed, it would not be justified to neglect 
dimension-ten operators, whose linear terms are actually less suppressed than the quadratic term
of a dimension-eight operator. And even worse, in the case of $s\sim \Lambda^2$ all higher-dimensional operators
could contribute equally and there is no physically meaningful
expansion anymore. 
This point simply denotes
the breakdown of the EFT approach. 
A related issue is the possible violation of unitarity, which
we will discuss in Sec.~\ref{sec:unitarity}.

In Fig.~\ref{fig:deltaRww_gg_sm_bsm} we display the observable $\Delta
R_{WW}=\sqrt{(y_1-y_2)^2+(\phi_1-\phi_2)^2}$, the separation in
rapidity and azimuthal angle between the two $W$ bosons.
Here we see that the effects of the higher-dimensional operators lead
to an enhancement of the distribution over the whole range.
Note that the region below $\Delta R_{WW}=\pi$ is not populated
because we only show the fixed-order results in this subsection.

In Figs.~\ref{fig:mww_gg_sm_bsm_8Tev} and \ref{fig:deltaRww_gg_sm_bsm_8Tev} we
show the same observables calculated at a center-of-mass energy of $\sqrt{s}=8$\,TeV.
We observe that the on-set of the BSM effects in the $m_{WW}$ distribution is around 550\,GeV. 
However, the relative size of the BSM contributions with respect to the SM contribution is much larger
at $\sqrt{s}=13$\,TeV,
as can be seen by comparing Figs.~\ref{fig:deltaRww_gg_sm_bsm} and~\ref{fig:deltaRww_gg_sm_bsm_8Tev}.
We have verified that the qualitative behaviour in comparing 13 to 8\,TeV stays the same 
if we also use a fixed scale ($M_{W}$) for the $\sqrt{s}=13$\,TeV case.
\begin{figure}[htb]
\centering
 \subfloat[\label{fig:mww_gg_scalevar}]{ \includegraphics[width=0.45\textwidth]{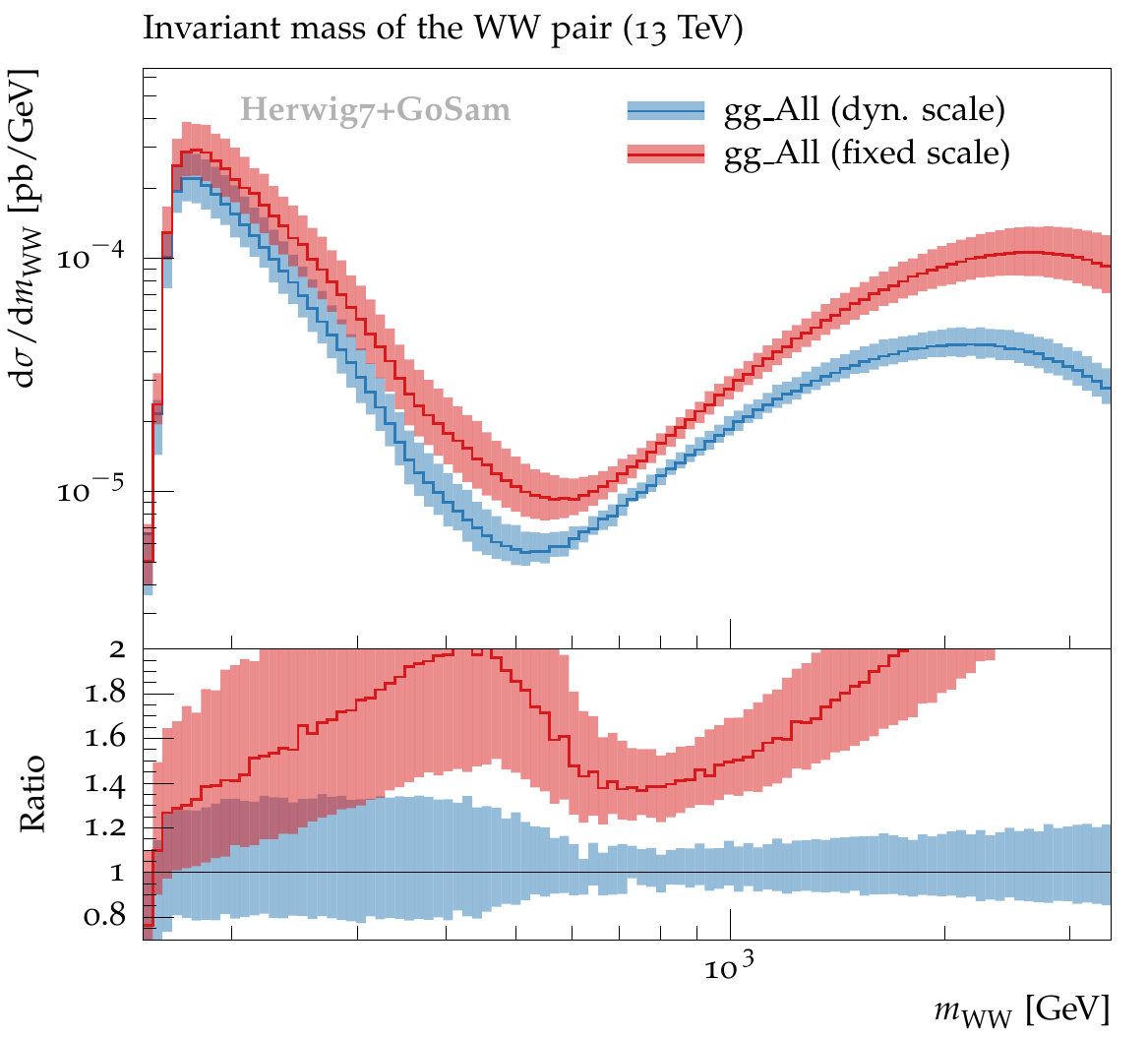}} \hfill
\subfloat[\label{fig:deltaRscalevar}]{ \includegraphics[width=0.45\textwidth]{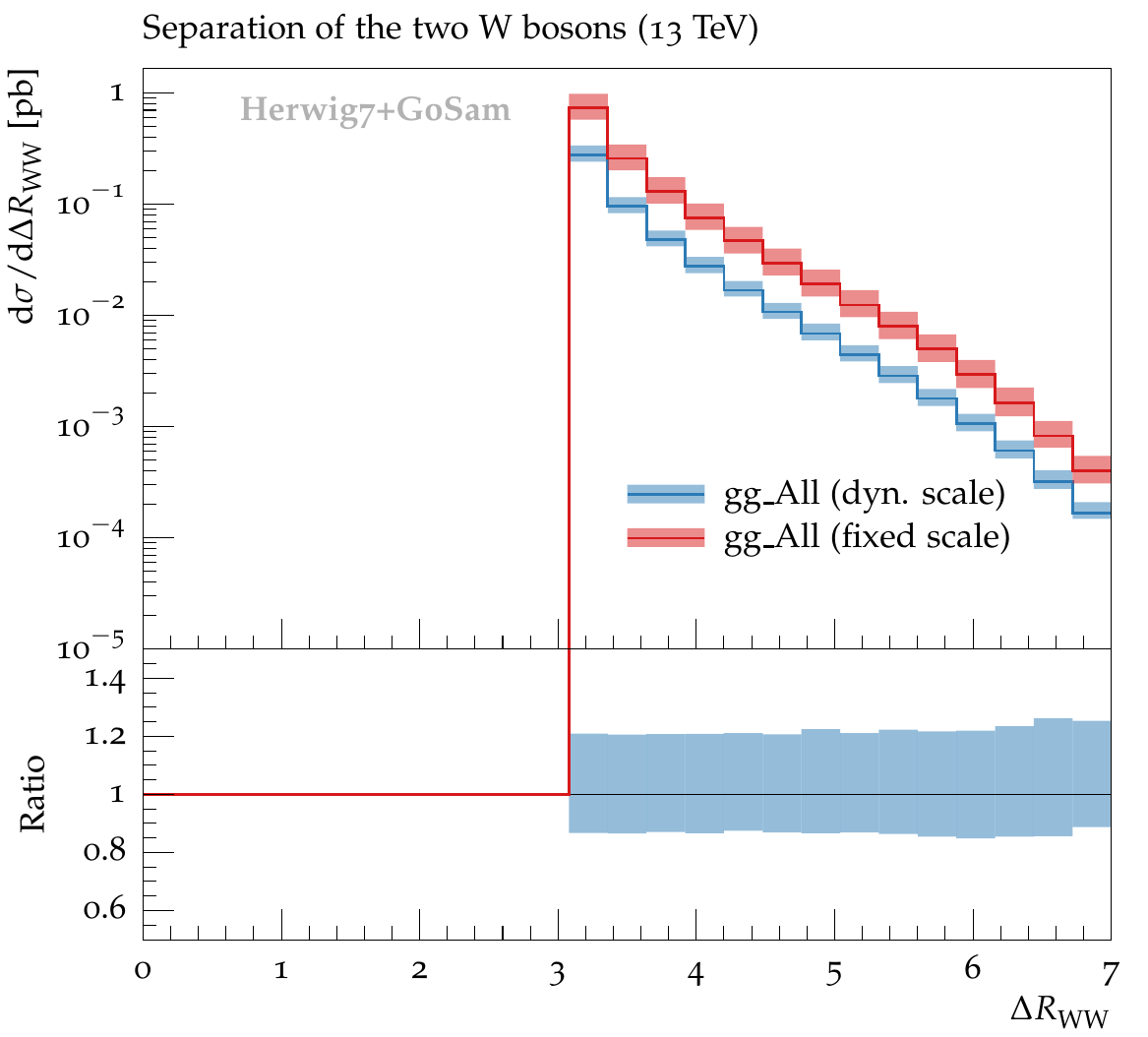}}
\caption{Scale variations, in the case of the $gg$-initiated contributions, for (a) the invariant-mass distribution of the
  $W$-boson pair and for (b) the $\Delta
  R_{WW}$ distribution.
The shaded bands show the scale-variation uncertainties. Ratio plots are with respect to \texttt{gg\_All\,(dyn.\,scale)}.
\label{fig:fixed_dyn}
} 
\end{figure}

In Fig.~\ref{fig:fixed_dyn} we compare the two scale choices 
$\mu_r=\mu_F=M_{W}$ (fixed) and 
$\mu_r=\mu_F=m_{WW}=\sqrt{(p_{e^+}+p_{\mu^-}+p_{\nu_e}+p_{\bar{\nu}_\mu})^2}$
(dynamic), including the scale uncertainty band,  
obtained as usual by varying by a factor of two up and down from those central scale choices.
The fixed scale $M_{W}$, being relatively low, leads to a larger value of $\alpha_s$ and therefore an increase in the cross section.
Since the bands do not overlap, this also means that the scale variations by factors of two up and down are not sufficient to capture the uncertainties correctly.
\begin{figure}[htb]
\centering
\subfloat[\label{fig:dphi_ll_gg}] {\includegraphics[width=0.45\textwidth]{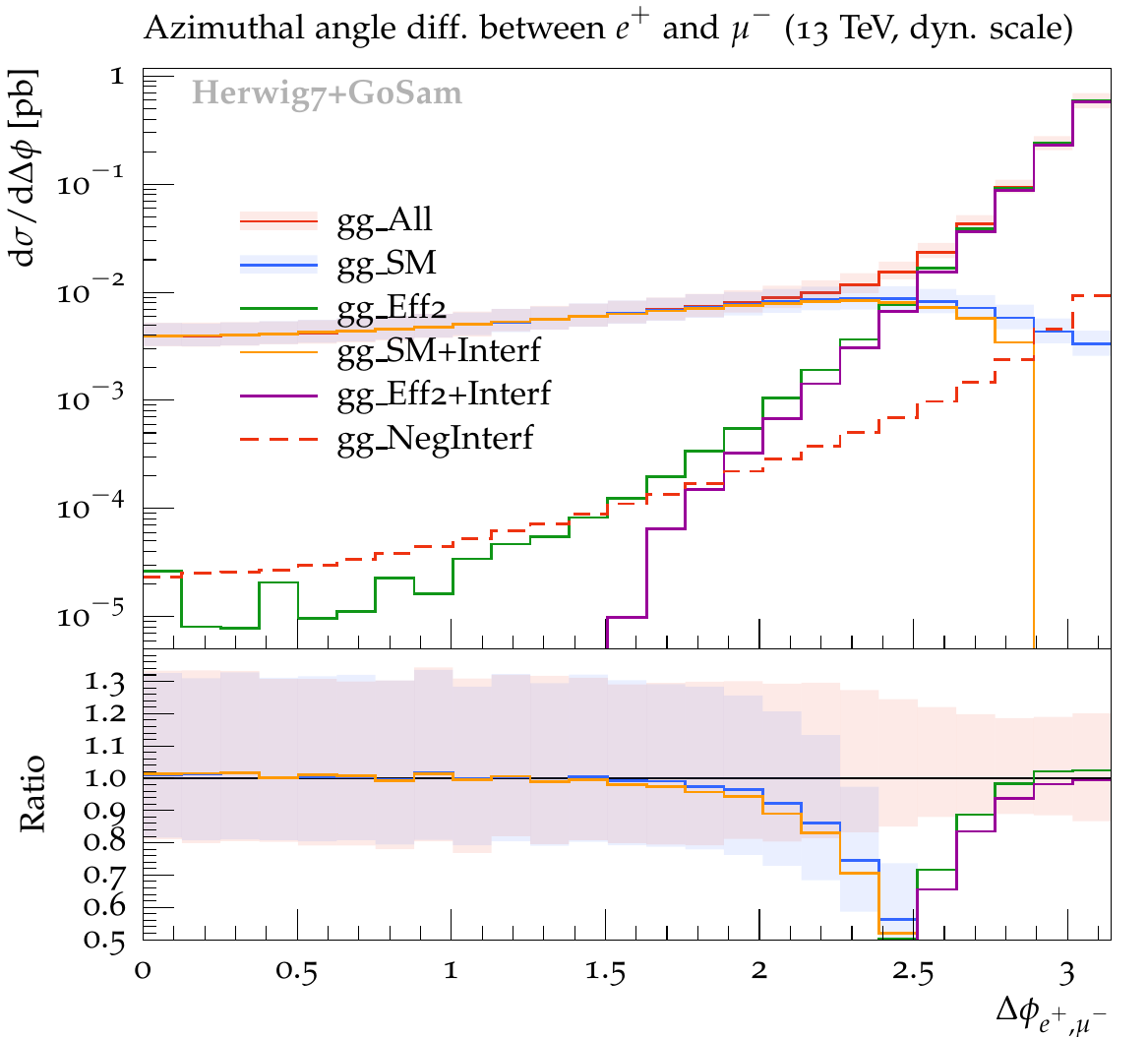}}\hfill
 \subfloat[\label{fig:deltaRll_gg}]{ \includegraphics[width=0.45\textwidth]{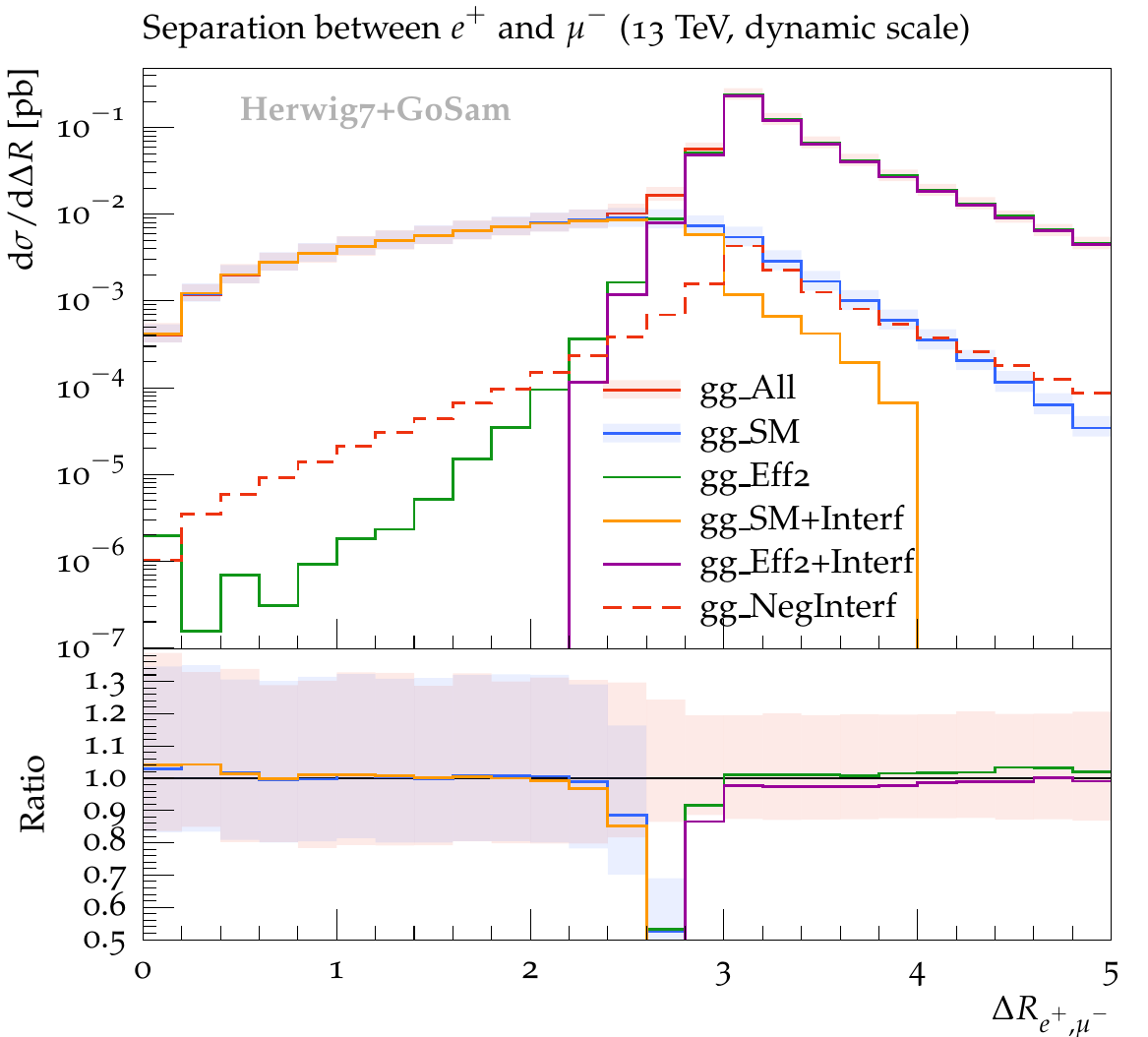}}
\caption{Distributions of (a) relative azimuthal angle $\Delta\phi_{e^+\mu^-}$
  (b) and $\Delta R_{e^+\mu^-}$,
for the $gg$-initiated contributions.
Ratio plots are with respect to \texttt{gg\_All}.
The shaded bands show the scale-variation uncertainties. 
}
\end{figure}
\begin{figure}[htb]
\centering
 \subfloat[\label{fig:ptw_gg}]{ \includegraphics[width=0.45\textwidth]{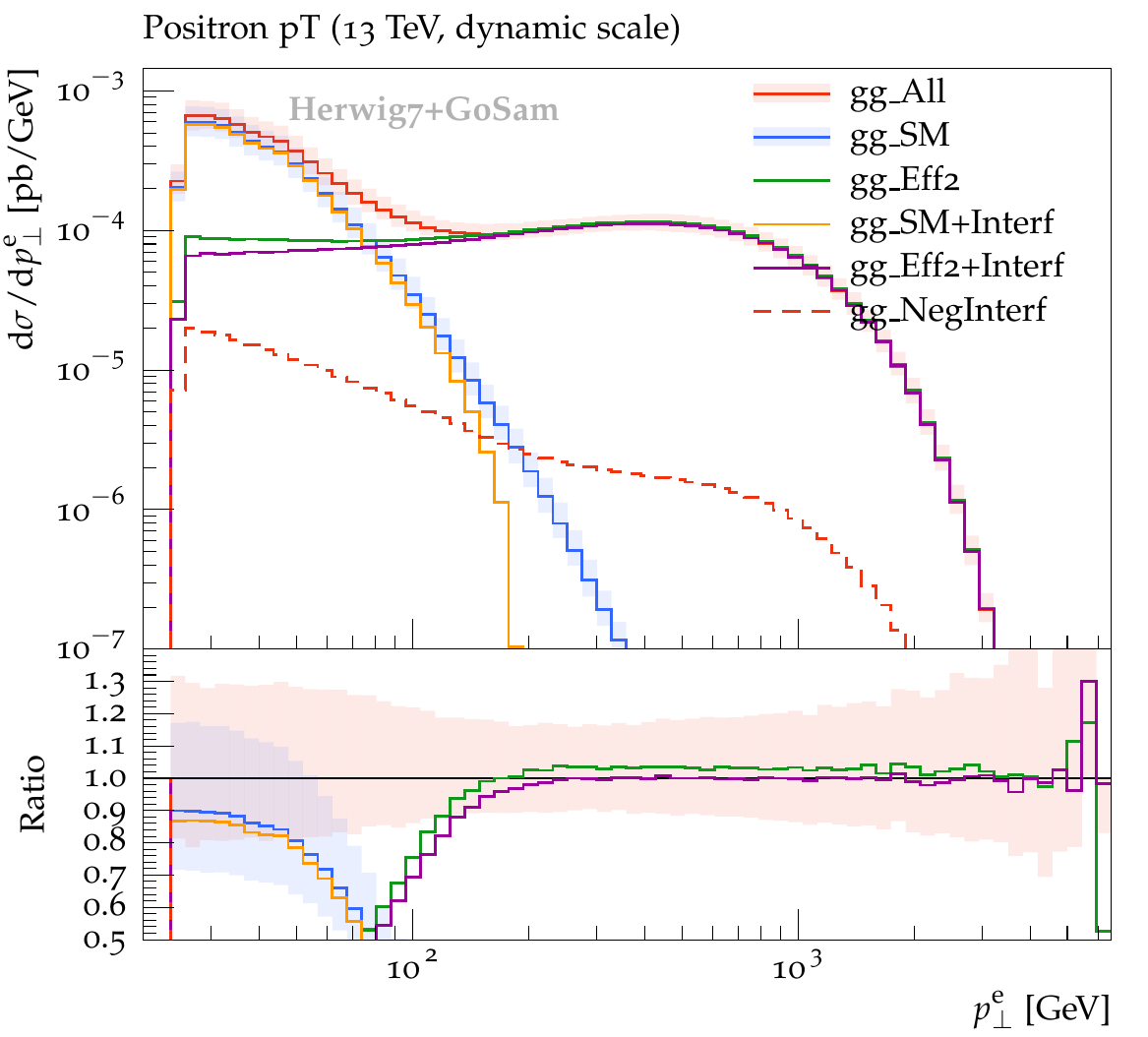}} \hfill
 \subfloat[\label{fig:mll_gg}]{ \includegraphics[width=0.45\textwidth]{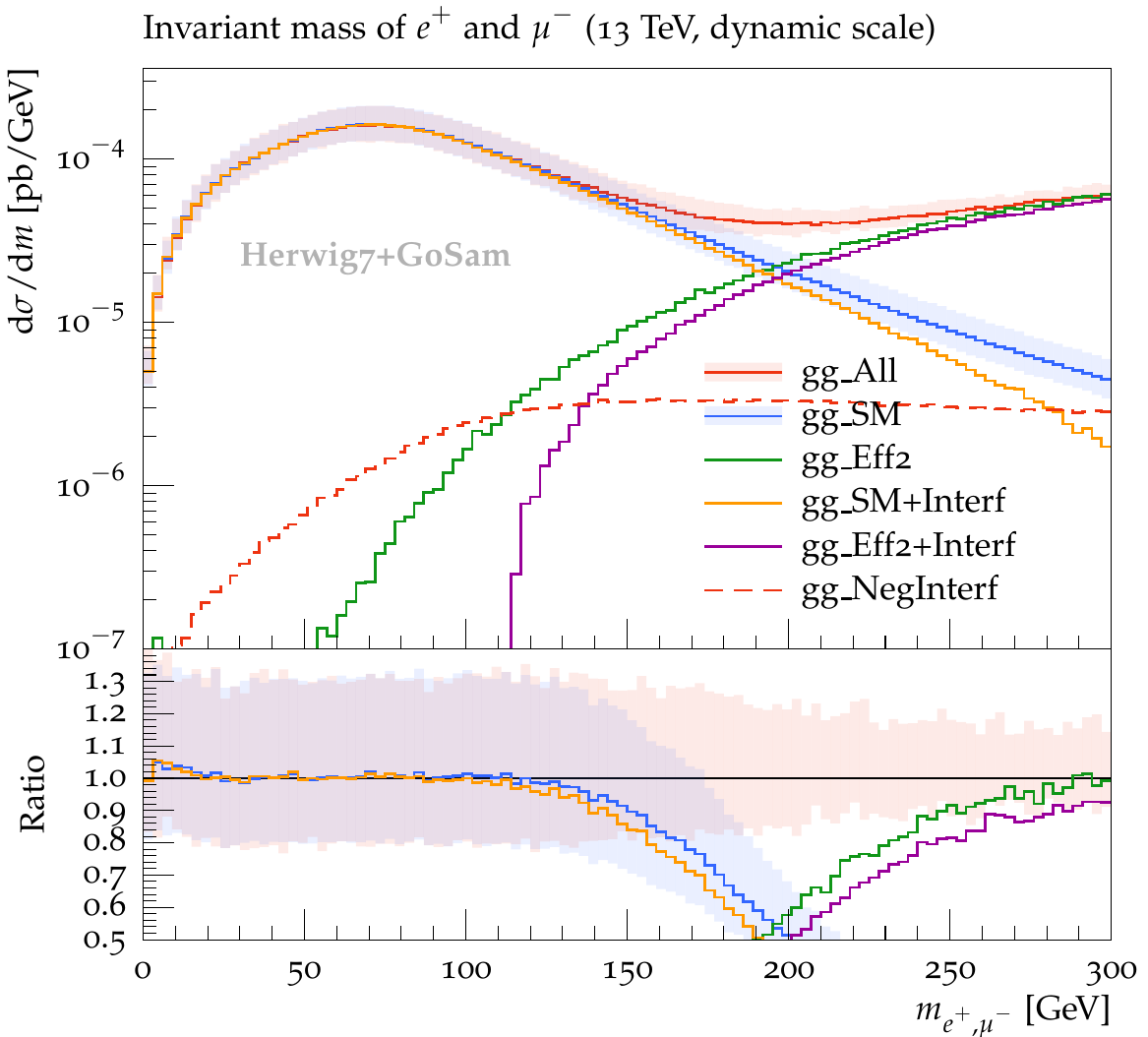}}
\caption{Transverse momentum of the positron 
(a) and  invariant mass of the charged leptons (b), for the $gg$-initiated contributions.
Ratio plots are with respect to \texttt{gg\_All}.
The shaded bands show the scale-variation uncertainties. 
} 
\end{figure}

Another interesting distribution is the relative azimuthal angle between the two charged leptons, 
$\Delta\phi_{e^+\mu^-}$, shown in Fig.~\ref{fig:dphi_ll_gg}. The contributions from the higher-dimensional operators lead to 
more highly boosted $W$ bosons, and therefore the associated leptons are more likely to be ``back-to-back''.
A similar behaviour can be seen in the $\Delta R$ distribution of the leptons (see Fig.~\ref{fig:deltaRll_gg}).

We also show the various contributions to the transverse momentum of the positron from the $W^+$ decay
in Fig.~\ref{fig:ptw_gg}  and the invariant mass of the charged leptons in Fig.~\ref{fig:mll_gg}.
The BSM effects lead to a clear enhancement in the $p_\perp^{e}$ spectrum, which becomes 
quite substantial already for $p_\perp^{e}$ values as low as $\sim 60\text{\,\,-\,}100$\,GeV. 
In the $m_{e^+\mu^-}$ distribution, the effect of the higher-dimensional operators is also clearly visible for energies 
larger than about 150\,GeV\,to\,190\,GeV already (taking scale-variation uncertainties into account).
However, as this concerns only the $gg$-initiated contribution, the effect will be washed out once 
all sub-processes contributing to the
$pp$ initial state, plus higher-order corrections, are taken into account, 
as will be discussed in Sec.~\ref{sec:pp}.

\medskip

\begin{figure}[t!]
\centering
\subfloat[\label{fig:WW_costheta_planes_c123comparison_13tev}] { \includegraphics[width=0.45\textwidth]{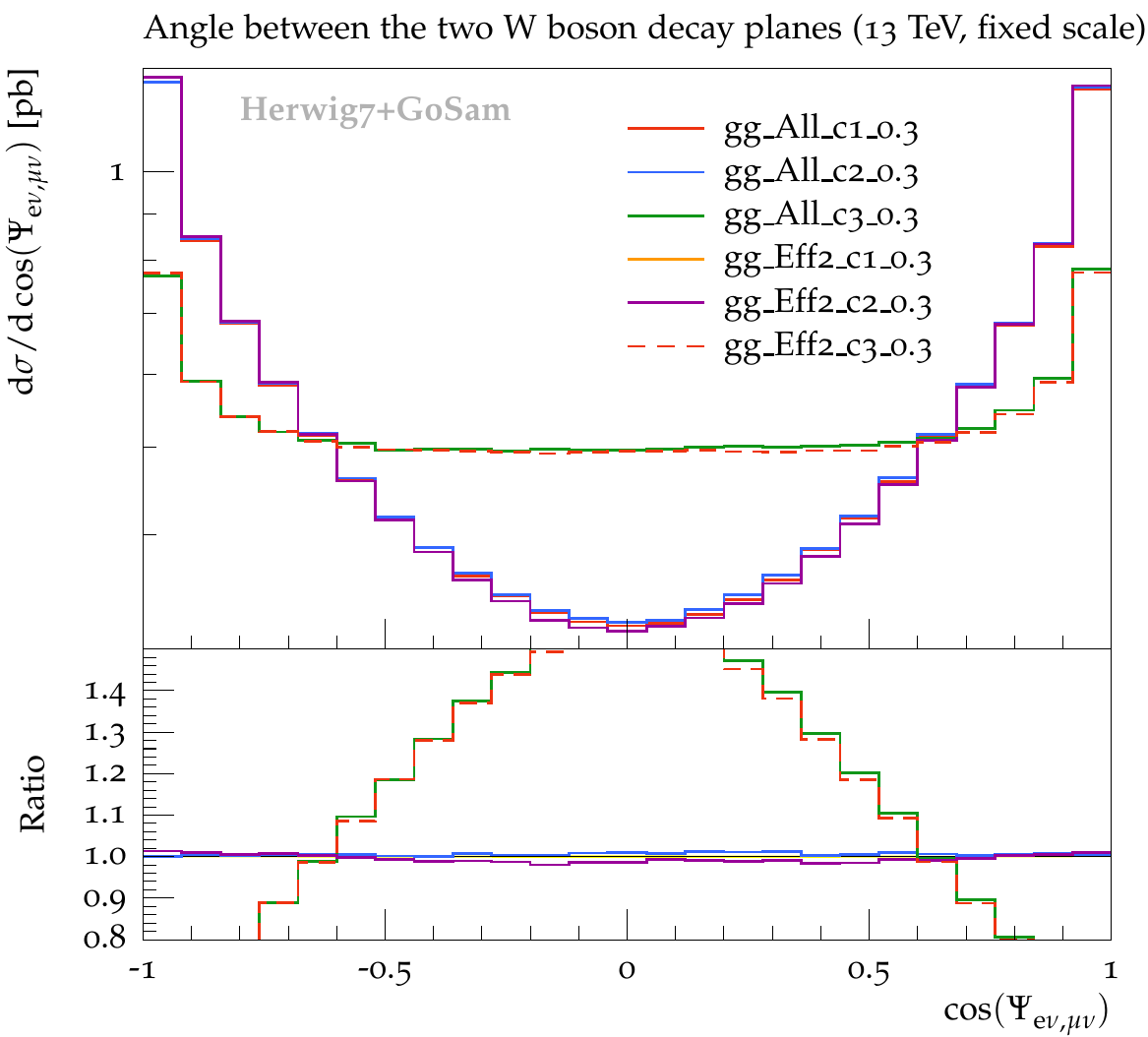}}\hfill
\subfloat[\label{fig:WW_costheta_planes_c123comparison_8tev}] { \includegraphics[width=0.45\textwidth]{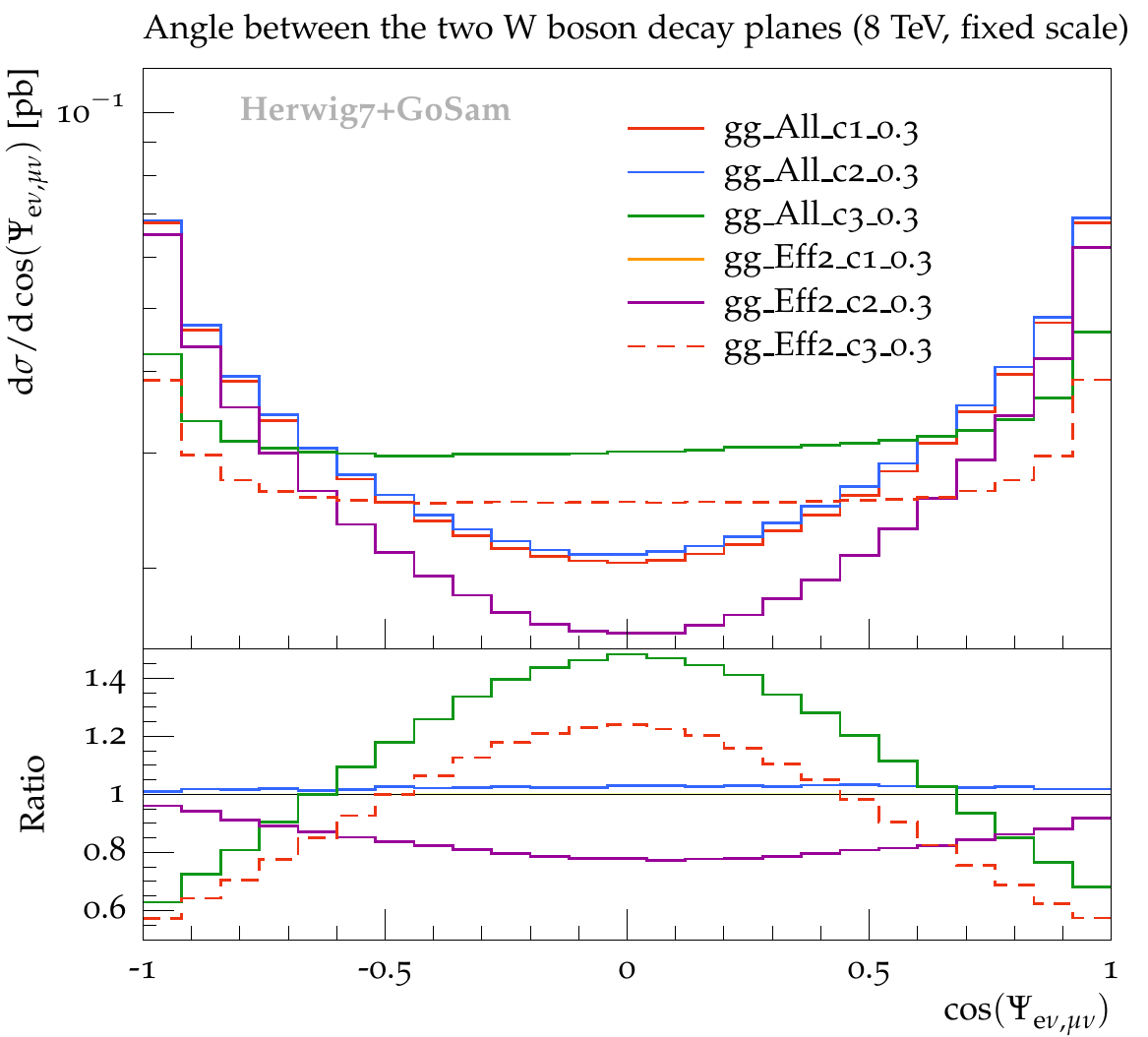}}
\caption{Angle between the $W$-boson decay planes for (a) 13\,TeV and (b)
  8\,TeV. At both energies a fixed scale of $\mu_r=\mu_F=M_W$ has been used.
Ratio plots are with respect to \texttt{gg\_All\_c1\_0.3} ($c_1=0.3,\,c_2=c_3=0$). 
Comparing (a) and (b) we see that the differences to the contributions from the third operator ${\cal O}_3$ are more prominent at 13\,TeV compared to 8\,TeV.
We also note that at 13\,TeV the linear term in ${\cal O}_3$ has a bigger effect compared to 8\,TeV.}
\label{fig:ci}
\end{figure}

\begin{figure}[t!]
\centering
\subfloat[\label{fig:MWW_ci_13tev}] {\includegraphics[width=0.45\textwidth]{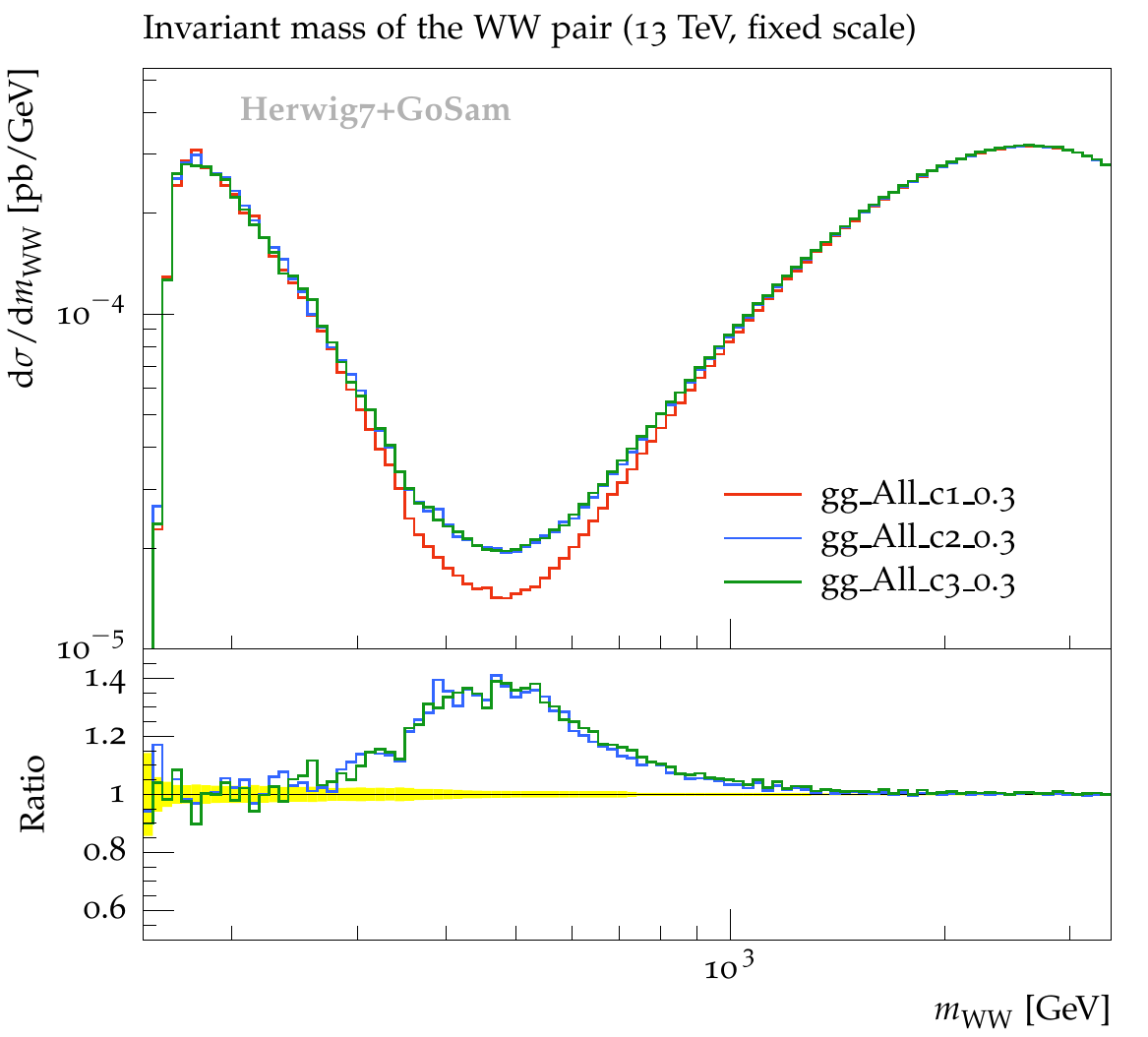}}\hfill
\subfloat[\label{fig:mww_c123comparison_13tev}] {\includegraphics[width=0.45\textwidth]{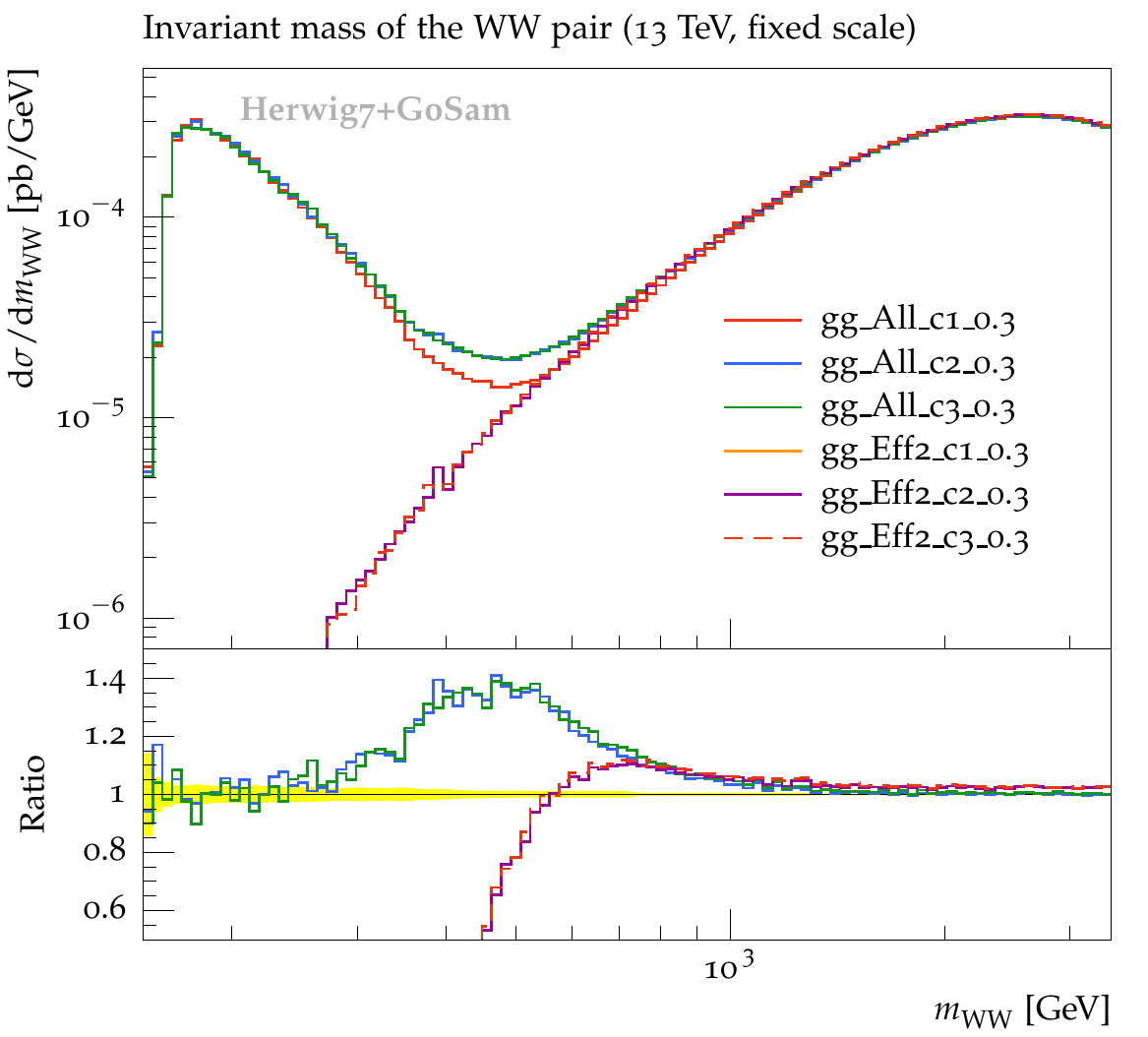}}
\caption{Invariant mass of the $W$-boson pair for various values of the anomalous couplings for the $gg$-initiated contributions, at 13 TeV.
(a) shows the sum of SM and anomalous contributions. (b) shows the same as (a) but with the pure squared EFT contributions shown in addition.
Note that the contributions from the first and second EFT operator, ${\cal O}_1$ and ${\cal O}_2$, 
are identical in the pure squared EFT contribution (orange and purple curves), 
while the interference terms involving ${\cal O}_1$ and ${\cal O}_2$, respectively, are different, as can be seen in the differences of  
the red and blue curves, which contain the interference terms.
Ratio plots are with respect to \texttt{gg\_All\_c1\_0.3} ($c_1=0.3,\,c_2=c_3=0$).
\label{fig:ciMww}}
\end{figure}

In order to investigate differences between the three dimension-eight operators, we will now consider them one at a time,
always setting the coupling constant of the two others  to zero respectively. In
Figs.~\ref{fig:ci} and \ref{fig:ciMww} the effects of the individual 
operators are shown for two observables, the angle between the decay planes of the $W$ bosons, $\cos{(\Psi_{e\nu,\mu\nu})}$,
and their invariant mass, $m_{WW}$. For these comparisons we have always set one of the $c_i$ 
coefficients to the value 0.3 and the other two to zero respectively.
Looking at the decay planes of the $W$ bosons in Fig.~\ref{fig:ci}, we find that the first two operators, ${\cal O}_1$ and ${\cal O}_2$, show the same angular
dependence, whereas the angular dependence of the third operator ${\cal O}_3$, 
which contains the dual field-strength tensor $\tilde{W}^{I,\mu\nu}$, is different,
which is seen to be more prominent for $\sqrt{s}=13$\,TeV.

To distinguish between
the first and the second operator, the invariant-mass distribution is also a suitable observable, as can be 
seen in Fig.~\ref{fig:MWW_ci_13tev}. Here the first operator leads to a stronger decrease of the distribution in 
the region around $m_{WW}\sim 500$\,GeV. Therefore, the combination of these two observables in principle allows 
for a distinction between the three operators. However, it should be noted that this can only be a qualitative
discussion, as the impact of the dimension-eight operators strongly depends on the size (and on the sign)
of the coupling constants $c_i$.
The overall size of the BSM effects for our default choice of the anomalous couplings is discussed in 
Sec.~\ref{sec:pp}, where we combine all sub-processes contributing to the $pp$ initial state.

\vspace{15pt}
\subsubsection{Unitarity bounds}
\label{sec:unitarity}
In the context of higher-dimensional operators it is also important to talk about unitarity. 
As the effects of these operators
grow with increasing center-of-mass energy, they will eventually violate unitarity. 
For the case of stable $W$ bosons, {\it i.e.} for $2\to 2$ scattering, a unitarity bound on the total cross section 
can be derived  along the lines of Ref.~\cite{Degrande:2012wf}. 
In more detail, we can start from Eq.~(48) of Ref.~\cite{Degrande:2012wf} 
(but use total angular momentum $J=0$  for the $gg$-initiated case),
where the bound for an inelastic $2\to 2$ scattering amplitude
$T^{\mathrm{\,in}}$, summed over the final state helicities
$\lambda_3,\lambda_4$, is given by
\be
\sum_{\lambda_3,\lambda_4}\int dPS_2\, |T^{\mathrm{\,in}}|^2 \; \leq \; 8\pi\;.
\label{ubound}
\ee
To obtain the bound for the total cross section, we include the flux
factor $ 1/(2\hat{s})$, average over initial state colours and
helicities, and sum over the colour and  helicity
configurations contributing to the cross section.
This leads to
\begin{eqnarray}
&&\sigma_{ggWW}\;=\;\frac{1}{2\hat{s}}\frac{1}{(N_c^2-1)^2}\frac{1}{4}\sum_{\mathrm{colours}}\,\sum_{\lambda_1,\lambda_2}\;
\underbrace{\sum_{\lambda_3,\lambda_4}\int dPS_2 |T^{\mathrm{\,in}}|^2 }_{\leq\,
  8\pi}\\
&\;\Leftrightarrow\;&\sigma_{ggWW}\;\leq\;
\frac{1}{2\hat{s}}  \sum_{\mathrm{colours}} \frac{\pi}{8}\;,
\end{eqnarray}
where we have used $\sum_{\lambda_1,\lambda_2\in\{+,-\}}= 4$.
The sum over the colour states contributing to the amplitude is given by
$\delta^{ab}\delta_{ab}=N_c^2-1$ (the trace of the identity matrix in the
adjoint representation).
Therefore, with $N_c=3$, we have
\be
\label{eq:xsec_limit}
\sigma_{ggWW}\;\leq\; \frac{\pi}{2\hat{s}}\;.
\ee

The derivation of the unitarity bound on the total cross section in Eq.~(\ref{eq:xsec_limit}) is based 
on a $2\to2$ scattering process. To obtain an estimate for the unitarity bound including the 
decay of the $W$ bosons one could integrate the $2\to 2$ process numerically and rescale the result
with the branching ratios of the two $W$ bosons decaying into leptons.
This procedure, however, does not take into account the effect of the cuts on the leptons, and therefore we refrain 
from showing a unitarity bound in the plots for the  distributions.

Unitarity arguments can also be employed  to calculate an upper bound on the absolute value of the anomalous
coupling constants. To do so we use the ansatz to require unitarity of the amplitude for a given set of
helicities and project the amplitude onto partial waves \cite{Jacob:1959at}. Looking at the scattering
$a+b\to c+d$ with the corresponding helicities $\lambda_{a},...,\lambda_{d}$, the partial wave decomposition reads
\begin{equation}
\label{eq:partial}
 \bra{\theta\phi\lambda_c\lambda_d}T(E)\ket{00\lambda_a\lambda_b} \;=\; 16\pi\sum_J(2J+1)\bra{\lambda_c,\lambda_d}T^J(E)
 \ket{\lambda_a,\lambda_b}e^{i(\lambda-\mu)\phi}d^J_{\lambda,\mu}(\theta)\;,
\end{equation}
with $\lambda=\lambda_a-\lambda_b$ and $\mu=\lambda_c-\lambda_d$, and where $\bra{\theta\phi\lambda_c\lambda_d}T(E)\ket{00\lambda_a\lambda_b}$
denotes the transition matrix element. Its unitarity must hold for each partial wave independently, {\it i.e.}
\begin{equation}
 |T^J|\;\le\; 1\;.
\end{equation}
Therefore we project the full amplitude onto single partial waves, where the strongest constraints typically
come from the lowest order partial waves. In the case where $\lambda=\mu=0$, the $d^J$ functions reduce to the Legendre polynomials,
{\it i.e.} $d^J_{00}(\theta)=P_J(\cos\theta)$.

Usually it is assumed that the strongest constraints stem from longitudinally polarized $W$ bosons, as in the limit
of large momentum $k$ the longitudinal polarization vector behaves like
\begin{equation}
 \lim_{k\to \infty} \epsilon^{\mu}_L(k) \;=\; \frac{k^{\mu}}{m}\;+\; {\cal{O}}\left(\frac{m}{E}\right)\;.
\end{equation}
Projecting onto the 0th partial wave we find
\begin{equation}
 \label{eq:limitlong}
 \left|\frac{c_1}{\Lambda^4}\right| \;,\; \left|\frac{c_2}{\Lambda^4}\right|\;\le\; \frac{2\pi}{M_W^2 \hat{s}}\;.
\end{equation}

For the third operator, ${\cal O}_3$, the contribution vanishes for longitudinally polarized $W$ bosons. It is also interesting 
to note that the contributions from the first two operators increase
more mildly with energy than naively expected. For dimensional
reasons the denominator in Eq.~(\ref{eq:limitlong}) could be $\sim s^2$, which in turn would mean that the amplitude itself 
could be $\sim s^2$. However we find only a behaviour which grows like $\sim s$.

The fact that the third operator vanishes for longitudinal polarizations, and  that we do not find the strongest
possible increase with the center-of-mass energy, suggests to also consider the situation where the 
$W$ bosons are transversely polarized. Projecting these amplitudes onto the 0th partial wave we find
\begin{equation}
 \label{eq:limittrans}
 \left|\frac{c_1}{\Lambda^4}\right| \;,\; \left|\frac{c_2}{\Lambda^4}\right|\;\le\; \frac{30\pi}{\hat{s}(26\hat{s}-11M_W^2)}\;,\quad 
 \left|\frac{c_3}{\Lambda^4}\right|\;\le\;\frac{\pi}{\hat{s}^{3/2}\sqrt{\hat{s}-M_W^2}}\;.
\end{equation}
This means that the strongest constraints for energies above the weak scale come from transversely polarized $W$ bosons
and, in order to maintain unitarity, one can roughly assume
\begin{equation}
 \left|\frac{c_i}{\Lambda^4}\right| \;\lesssim\; \frac{\pi}{\hat{s}^2}\;.
\end{equation}


\vspace{15pt}
\subsection{Impact of heavy-quark loop contributions}

We have taken both bottom- and top-quark masses into account for the
quark loops mediating the SM $gg\to W^+W^-$ interaction. 
The effect of the bottom-quark mass is negligible (of the order of the
Monte Carlo integration error), while  top-quark mass effects
have a considerable impact on the partonic cross section in the $gg$-initiated channel.

\begin{figure}[htb]
\centering
\subfloat[\label{fig:Memu_massless}] { \includegraphics[width=0.45\textwidth]{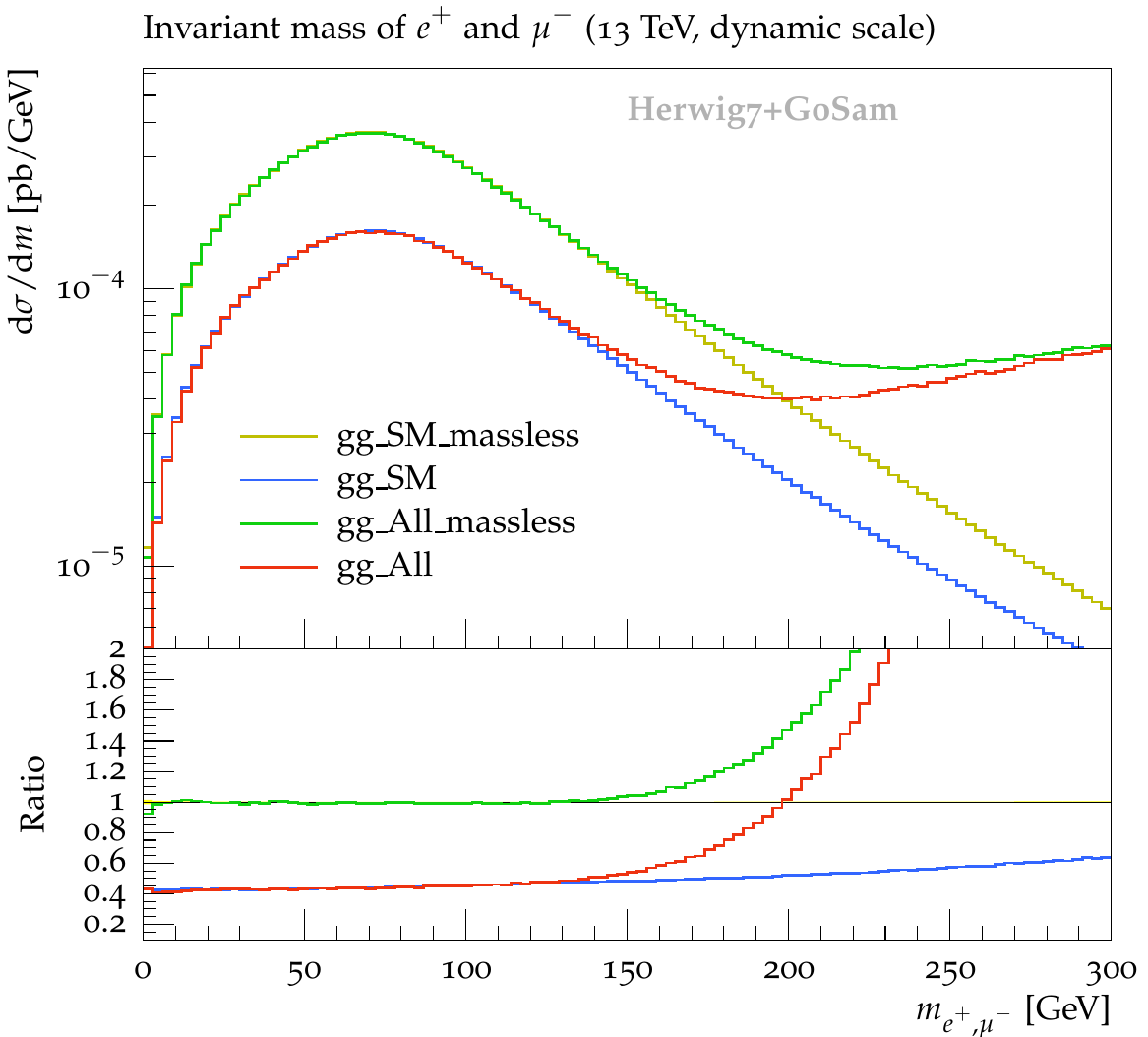}}\hfill
\subfloat[\label{fig:dRemu_massless}] { \includegraphics[width=0.45\textwidth]{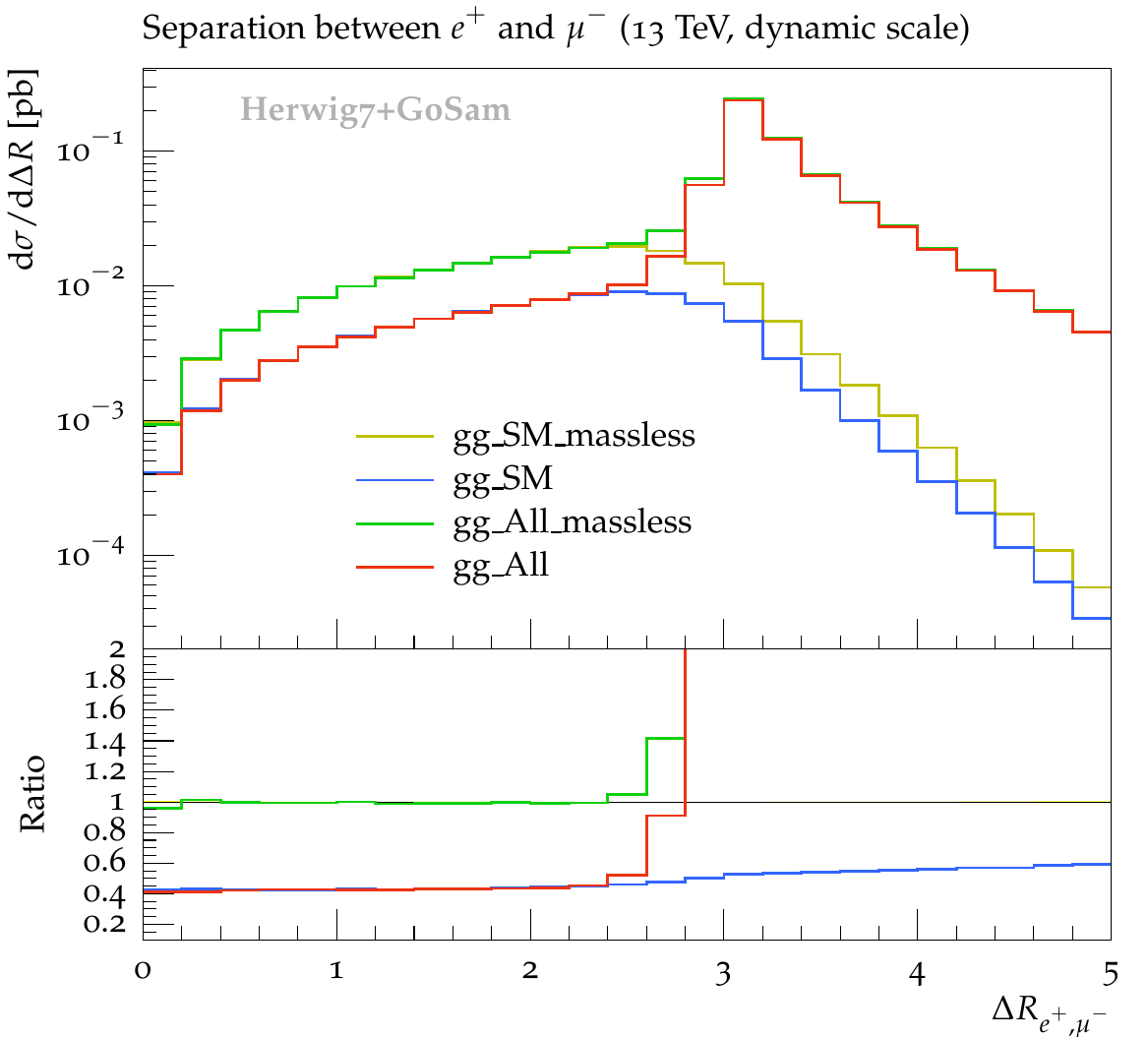}}\hfill
\caption{Top-quark mass effects on (a) the invariant mass and (b) the
  $R$-separation of the charged leptons. The curves labeled \texttt{\_massless} show results
for which the masses of the top and bottom quarks 
in the loops have been set to zero. 
Ratio plots are with respect to \texttt{gg\_SM\_massless.}
The range in $m_{e^+\mu^-}$ has been limited to 300\,GeV here for better visibility of the 
SM/BSM transition region. 
We have verified that for very large values of $m_{e^+\mu^-}$,  
the yellow and blue curves merge again, as expected.
\label{fig:topmasseffects}}
\end{figure}

Fig.~\ref{fig:topmasseffects} shows the effect of massive top-quark loops on the 
invariant-mass distribution of the charged leptons and on the
$R$-separation between the charged leptons.
We observe that top-mass effects decrease the $m_{e^+\mu^-}$
distribution by more than 30\% below values of $m_{e^+\mu^-}\sim250\text{\,-\,}300$\,GeV. The SM contribution with $M_t$ set to zero ({\tt  gg\_SM\_massless})
is of the same size as the SM+BSM result with masses taken into account ({\tt
  gg\_All}) at $m_{e^+\mu^-} \sim 200$\,GeV, which means that neglecting the
top-quark mass in the SM calculation could potentially ``fake'' BSM effects.

It should be noted here that in the case of massless top quarks also the Yukawa coupling between the Higgs boson and the top quark vanishes.

\begin{figure}[htb]
\centering
\subfloat[\label{fig:Memu_mt}] { \includegraphics[width=0.45\textwidth]{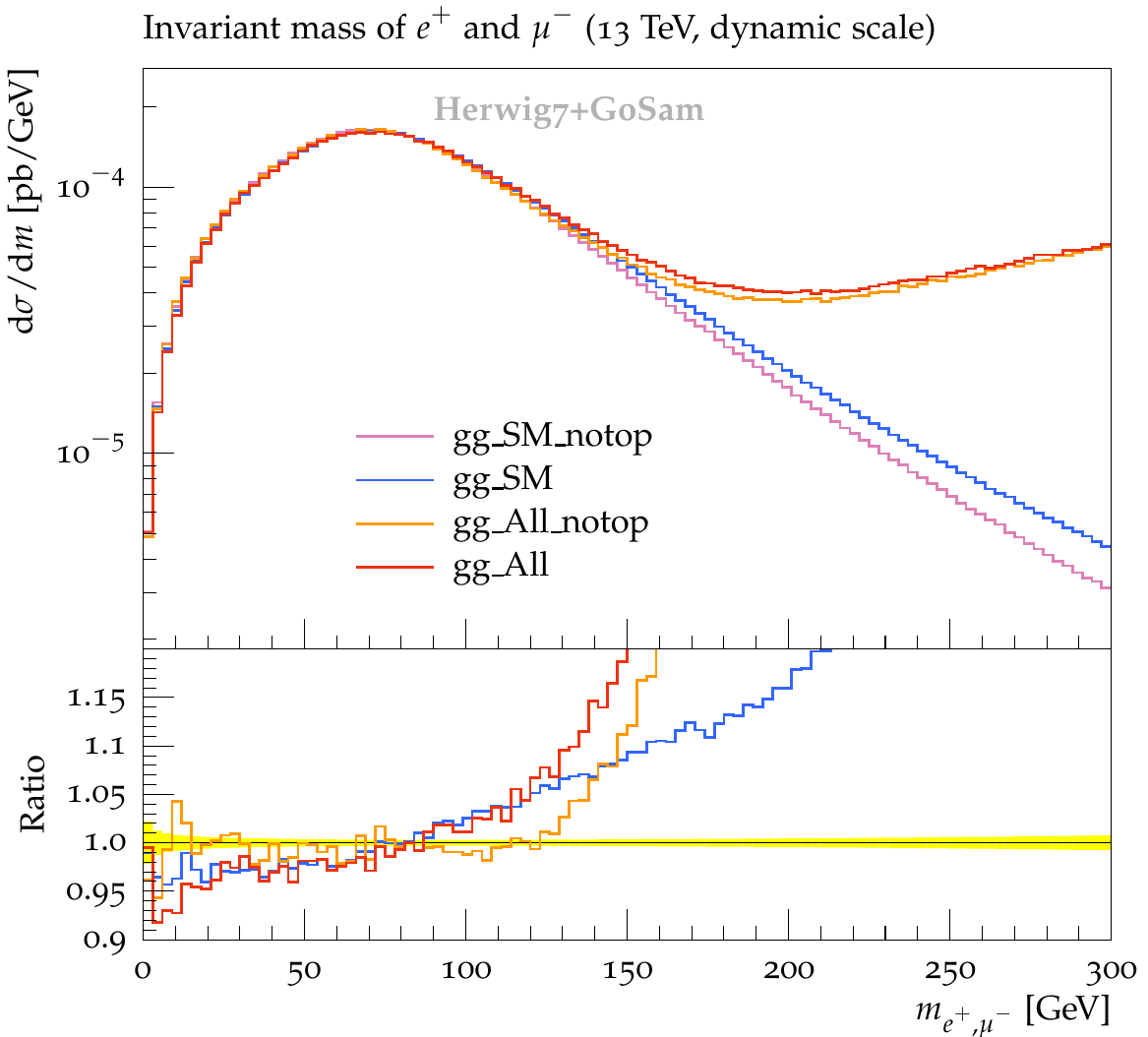}}\hfill
\subfloat[\label{fig:dRemu_mt}] { \includegraphics[width=0.45\textwidth]{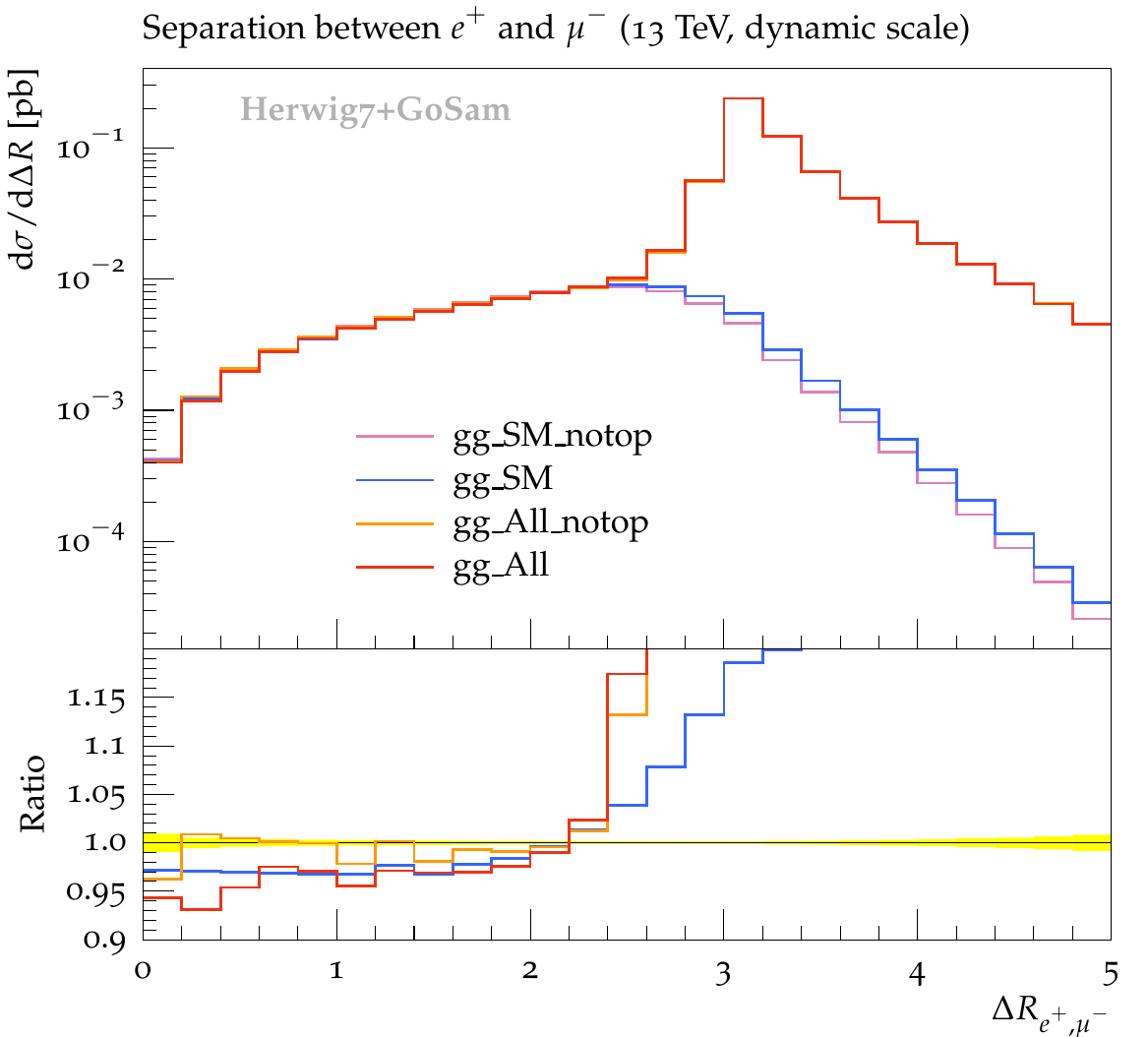}}
\caption{Impact of the top-quark loops on the invariant mass and the
  $R$-separation of the charged leptons.
The curves labeled \texttt{\_notop} show results for which diagrams with top-quarks in the loops have been omitted altogether.
Ratio plots are with respect to \texttt{gg\_SM\_notop.}\label{fig:toploopeffects}}
\end{figure}

\begin{figure}[htb]
\centering
\subfloat[\label{fig:M_ww_pp_topmass}] { \includegraphics[width=0.45\textwidth]{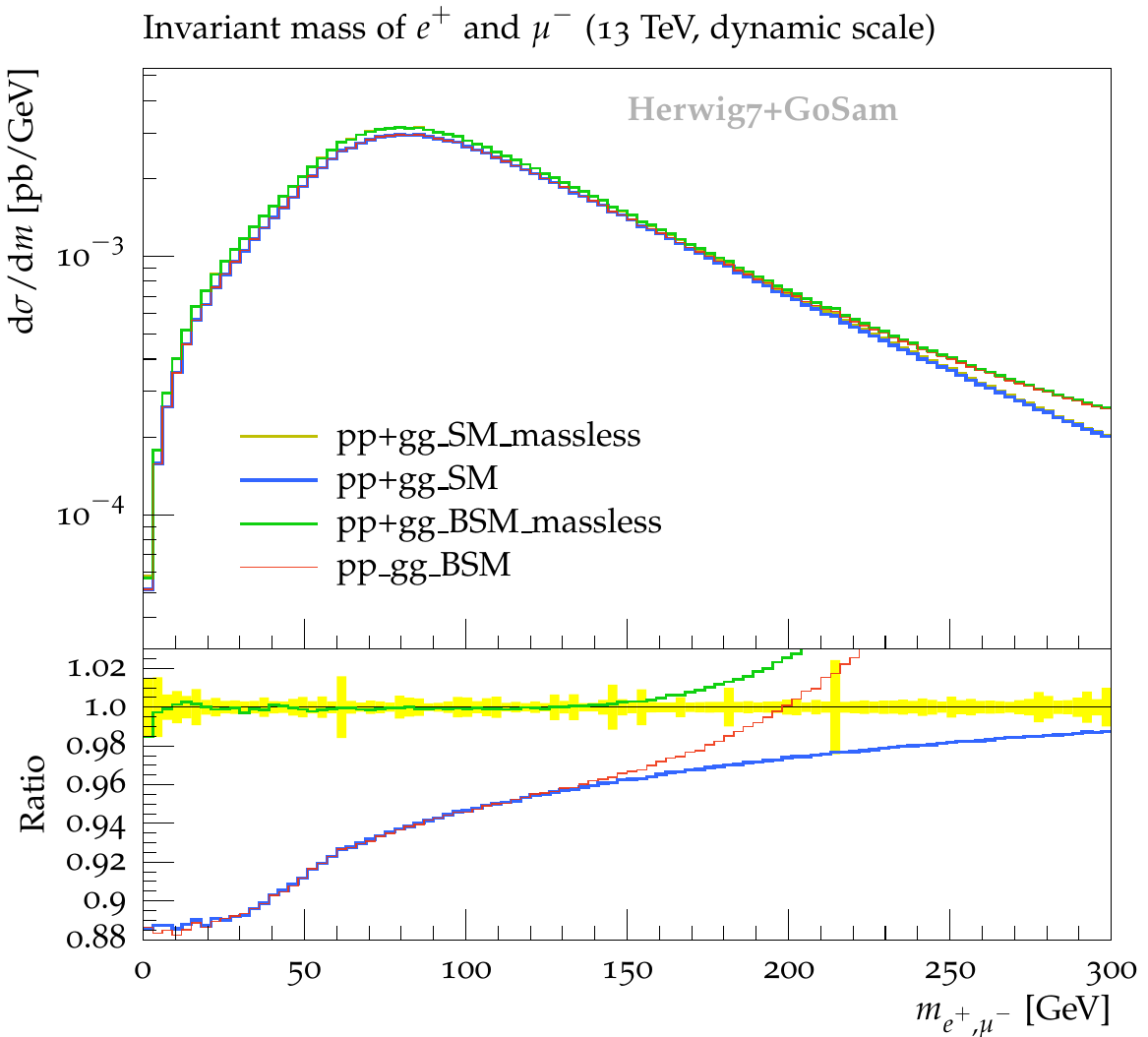}}
\subfloat[\label{fig:DeltaRww_pp_topmass}] { \includegraphics[width=0.45\textwidth]{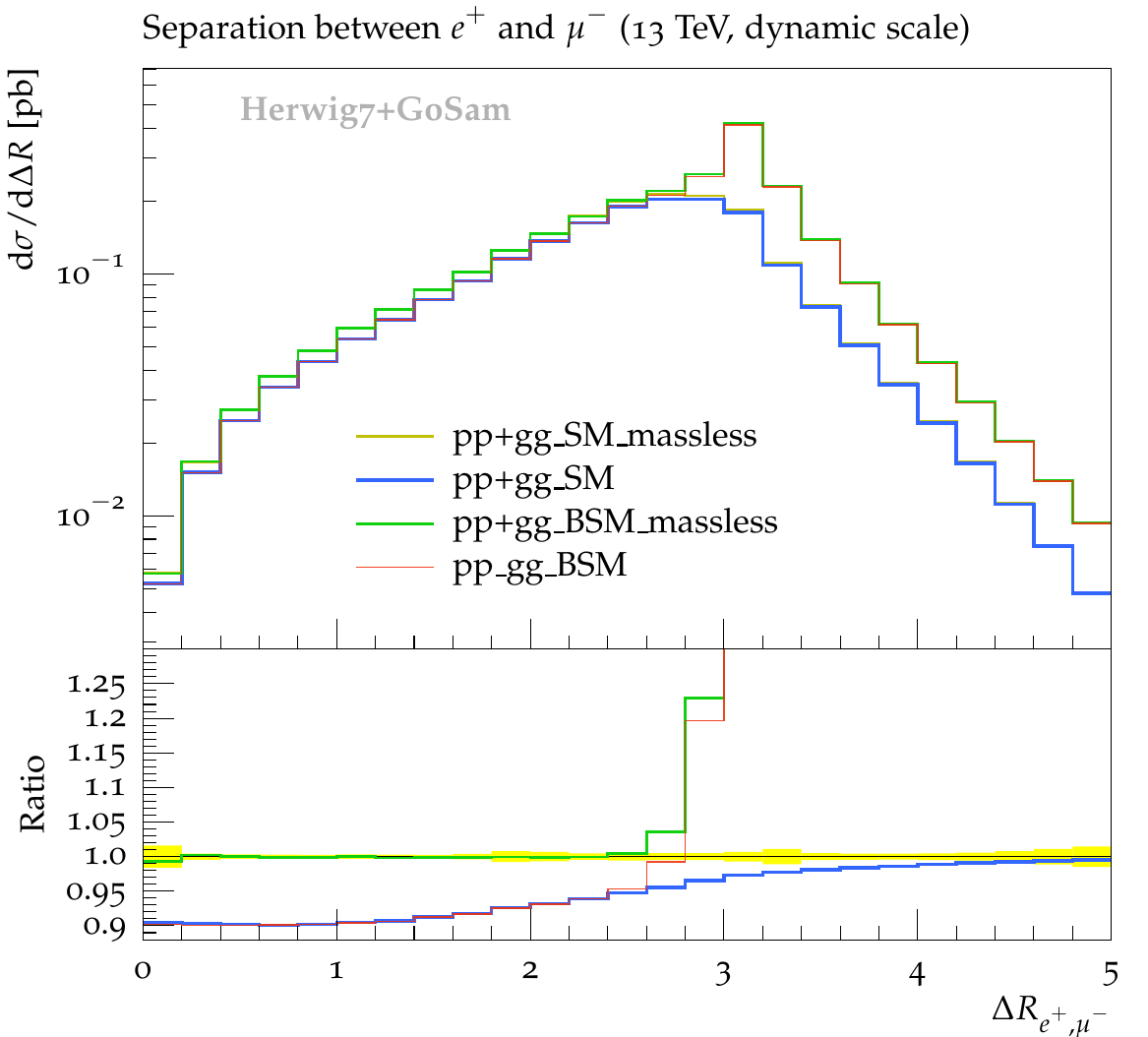}}
\caption{Impact of the top-quark loops on (a) the invariant mass of
  the charged leptons and (b) $\Delta R_{e^+\mu^-}$,  
 considering the full $pp$ initial state. Ratio plots are with respect to \texttt{pp+gg\_SM\_massless.}
\label{fig:pp_masseffects}}
\end{figure}

This result is contrasted to a calculation where the top-quark loops have been dropped
 altogether\footnote{It should be noted here that omitting the top quarks 
 also eliminates almost all contributions involving bottom quarks. 
 Only the diagrams where a Higgs boson couples to a b-quark pair remain, which are numerically 
 negligible. Therefore, omitting the top quark loops basically means excluding the third quark generation.}, 
shown in Fig.~\ref{fig:toploopeffects}. 
This has a considerable impact on the $m_{e^+\mu^-}$ distribution
beyond about 150\,GeV, however, the effect is much less pronounced than in
the case where top-quark loops are taken into account but the
top-quark mass is neglected ({\it cf.} Fig.~\ref{fig:topmasseffects}).
Even though the mass effects are below the 10\% level once the full $pp$ initial state including the NLO corrections 
is  taken into account (see Fig.~\ref{fig:pp_masseffects}), this study  demonstrates that 
massive top-quark loops should be taken into account to describe
measurements of highly  boosted $W$ bosons.

\vspace{15pt}
\subsection{Combination of gluon- and quark-initiated channels}
\label{sec:pp}

In this section we compare the $gg$-initiated contribution to the full process $pp\;(\to W^+W^-)\to e^+ \nu_e
\mu^-\bar{\nu}_\mu$ at NLO QCD,  considering first the results at the fixed-order level.
Shower effects will be discussed in Sec.~\ref{sec:ps}. The important questions here are how visible the effects of the 
anomalous couplings are if the quark-initiated Standard-Model contributions are added, and to what extent the 
effects of the higher-dimensional operators are hidden in the theoretical uncertainties.
\begin{figure}[htb]
\centering
\subfloat[\label{fig:WW_m_pp_scalevar_effects}]{\includegraphics[width=0.45\textwidth]{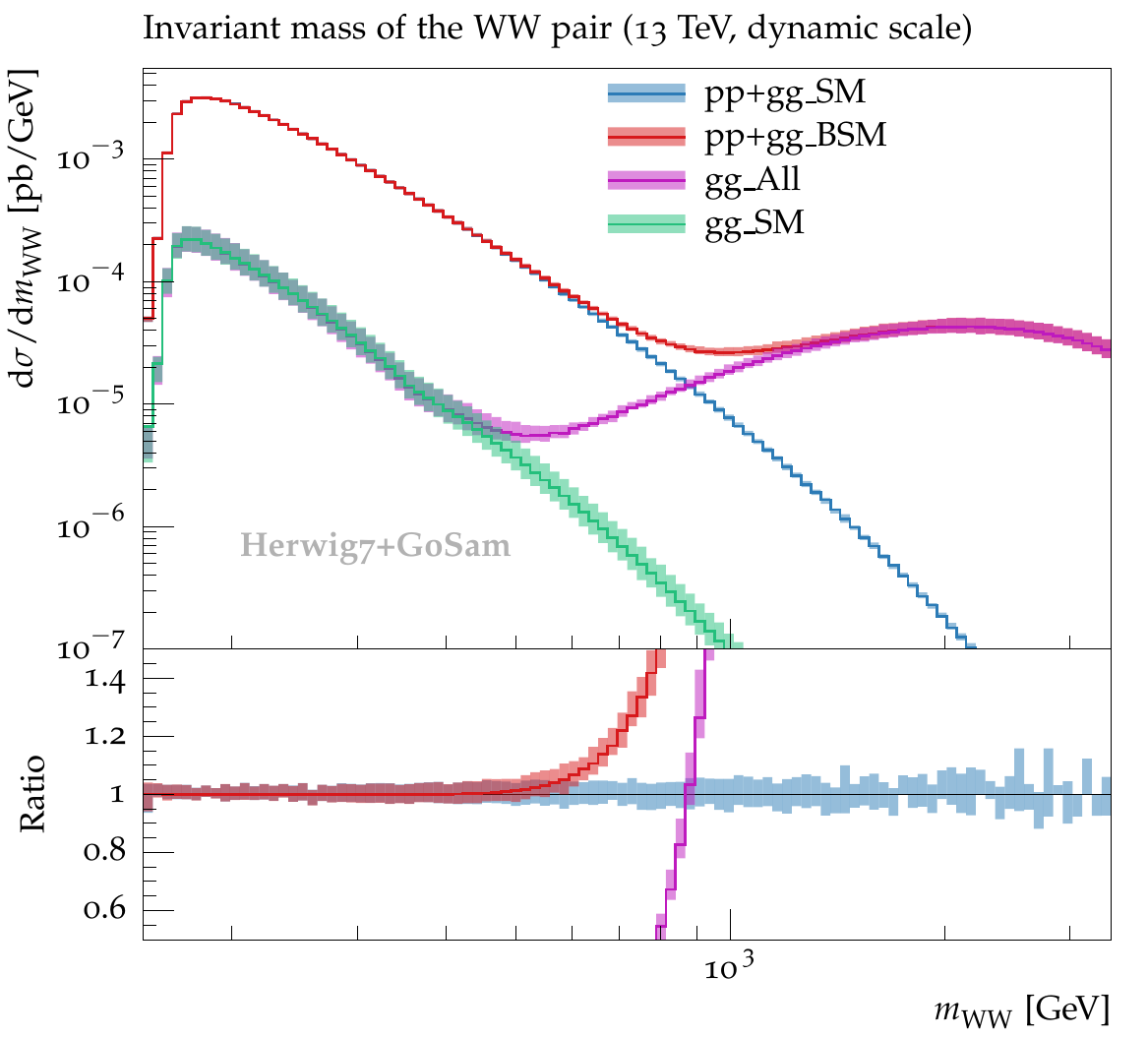}} \hfill
\subfloat[\label{fig:WW_dR_pp_scalevar_effects}]{\includegraphics[width=0.45\textwidth]{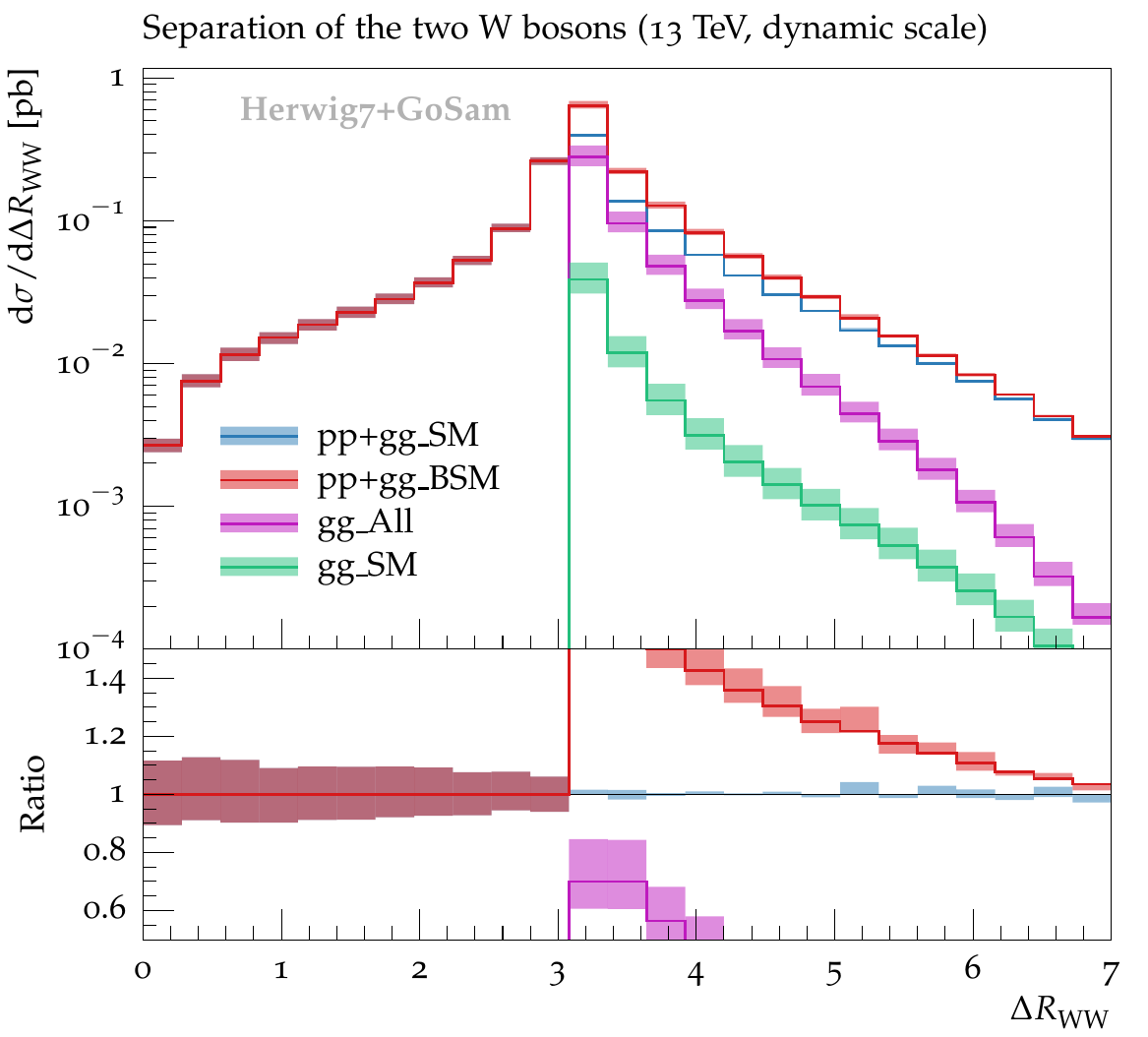}}
\caption{
Distributions for (a) invariant mass $m_{WW}$ and (b) separation $\Delta R_{WW}$ of the reconstructed $W$-boson pair
in $pp\;(\to W^+W^-)\to e^+ \nu_e\mu^-\bar{\nu}_\mu$ for the sum of all partonic channels,
including the effects of the higher-dimensional operators.
\texttt{pp+gg\_SM} includes all partonic channels, {\it i.e.} all the quark-initiated channels up to NLO QCD plus the loop-induced SM $gg$-initiated contribution \texttt{gg\_SM}.
\texttt{pp+gg\_BSM} is the same but includes the loop-induced SM+BSM $gg$-initiated contribution \texttt{gg\_All} instead of just \texttt{gg\_SM}.
In addition we show the SM and SM+BSM $gg$-initiated contributions separately.
The shaded bands show the scale-variation uncertainties.
Ratio plots are with respect to \texttt{pp+gg\_SM}.
}
\label{fig:mww_pp_scalevar}
\end{figure}

In Fig.~\ref{fig:mww_pp_scalevar}
we show the invariant-mass distribution of the $W$-boson pair including all SM as well as EFT contributions. 
The effects of scale variations are plotted as well,
where the scale uncertainty bands have been obtained by varying by a factor of two up and down from the dynamic scale choice $\mu_R=\mu_F=m_{WW}$.
We show the  SM NLO contribution with and without the EFT contributions,
and in comparison to that the effects of the higher-dimensional operators in the $gg$-initiated contributions alone.
This allows to directly assess the impact of the anomalous couplings. The loop-induced, $gg$-initiated Standard-Model
contribution leads to an ${\cal{O}}(10\%)$ increase over the quark- or quark-gluon-initiated NLO result. 
We therefore observe that in the full result, combining all channels,
a visible deviation from the SM prediction begins to show at larger $m_{WW}$ values, of about 700\,GeV, while 
in the $gg$-initiated contribution, shown in Fig.~\ref{fig:mww_gg_sm_bsm}, the deviation already starts 
to be visible at about 500\,GeV\,to\,600\,GeV (taking scale-variation uncertainties into account). 
While for  values of $m_{WW}$ around 700\,GeV the size of the BSM effects is comparable to the size of the scale uncertainties, 
shown in Fig.~\ref{fig:mww_pp_scalevar}, 
for values of $m_{WW}$ of about 800\,-\,900\,GeV,
the deviations from the Standard Model
due to the higher-dimensional operators start to become 
clearly 
visible.
On the other hand, the region beyond 1 TeV already probes energies where the EFT approach starts to become invalid.

\vspace{15pt}
\subsection{Parton-shower effects}
\label{sec:ps}

\begin{figure}[htb]
\centering
\subfloat[\label{fig:WW_dR}]{ \includegraphics[width=0.45\textwidth]{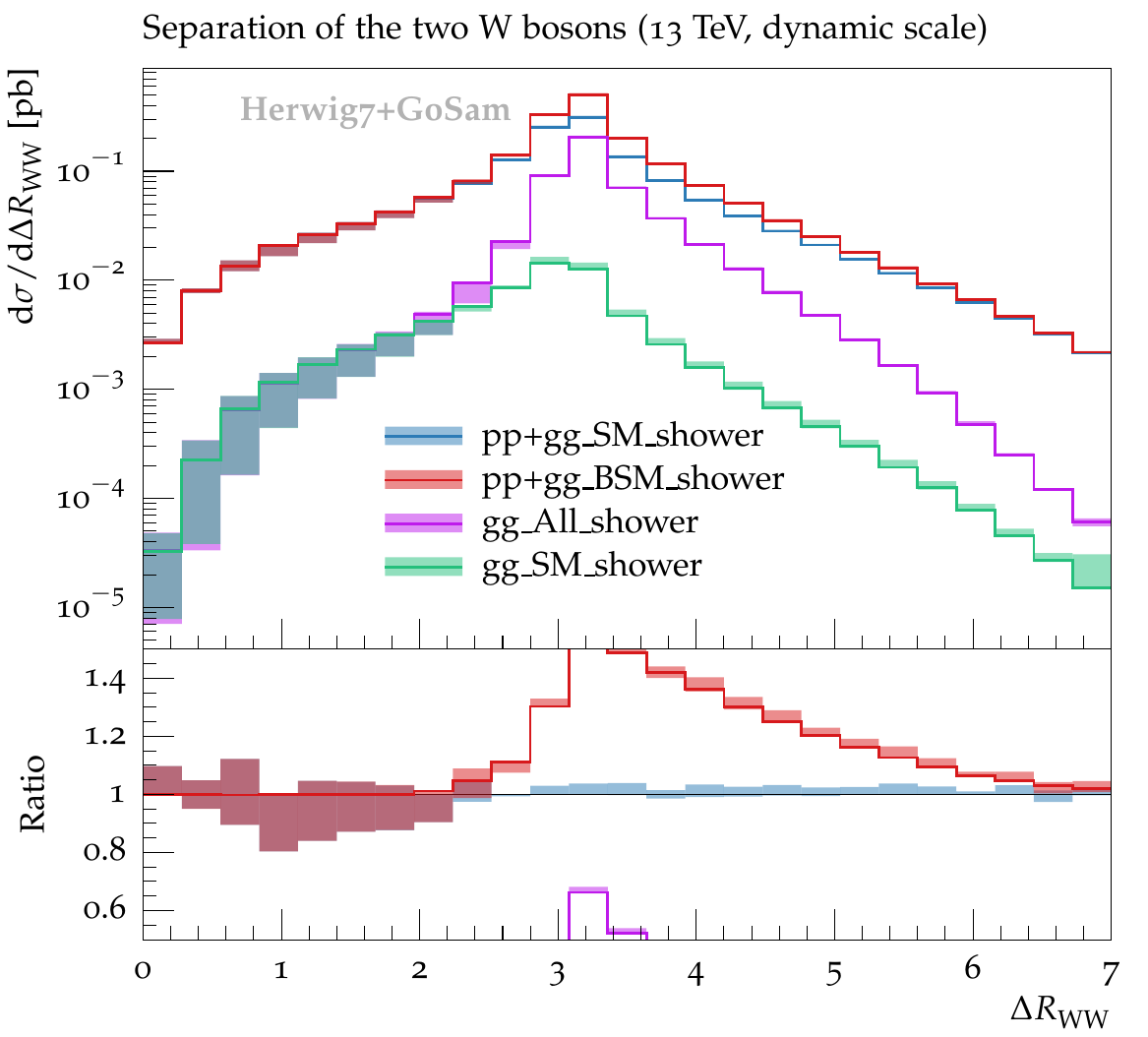}} \hfill
\subfloat[\label{fig:WW_pT}]{ \includegraphics[width=0.45\textwidth]{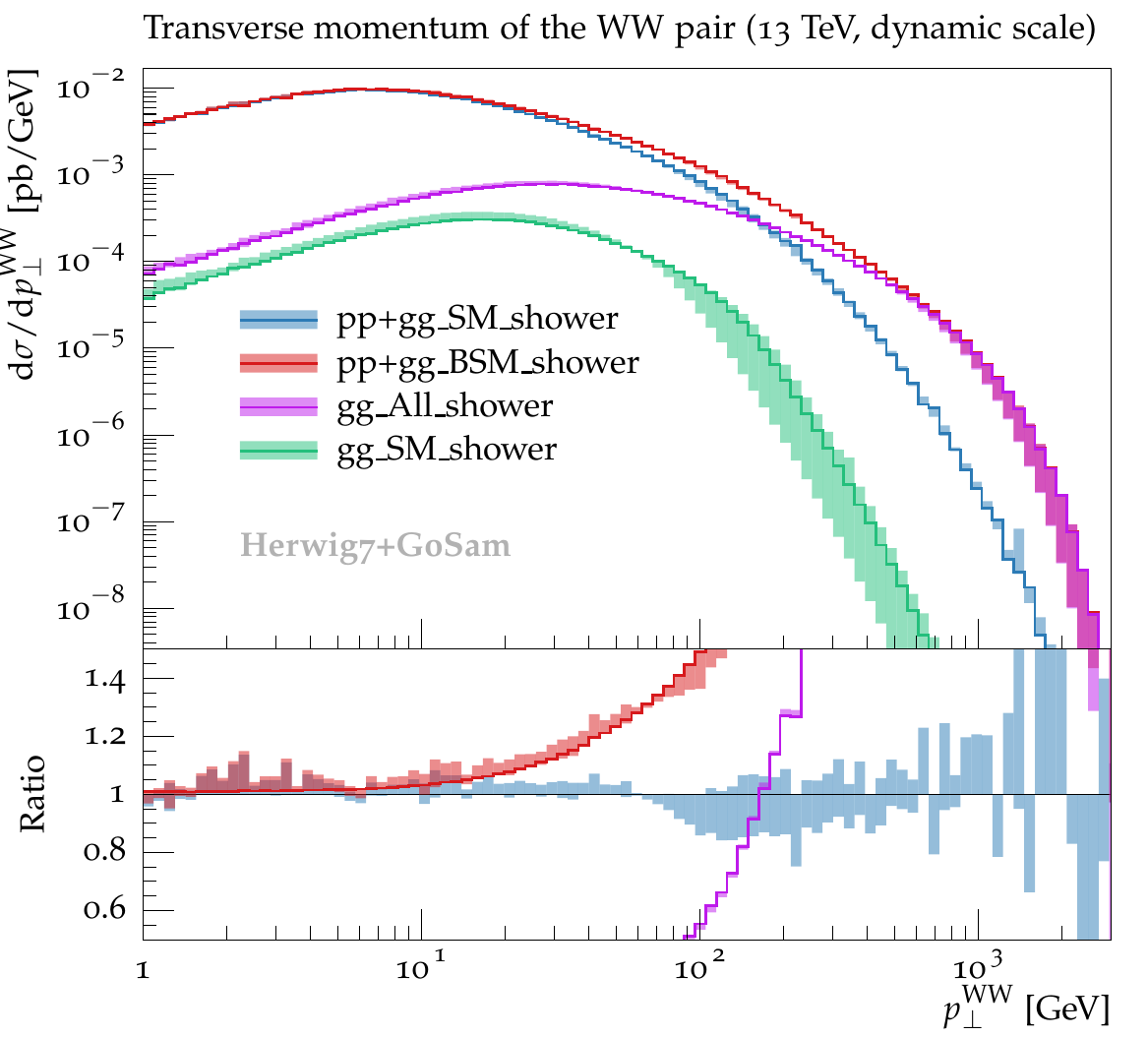}}
\caption{$\Delta R_{WW}$ and $p_{\perp}^{WW}$ distributions for the sum of all
  partonic channels contributing to $pp\;(\to W^+W^-)\to e^+ \nu_e\mu^-\bar{\nu}_\mu$,
  including $\mu_Q$ variations and effects of the higher-dimensional operators.}\label{fig:IR_sensitive}
\end{figure}

\begin{figure}[htb]
\centering
\subfloat[\label{fig:eWmuW_dR_pp_muQvar}]{ \includegraphics[width=0.45\textwidth]{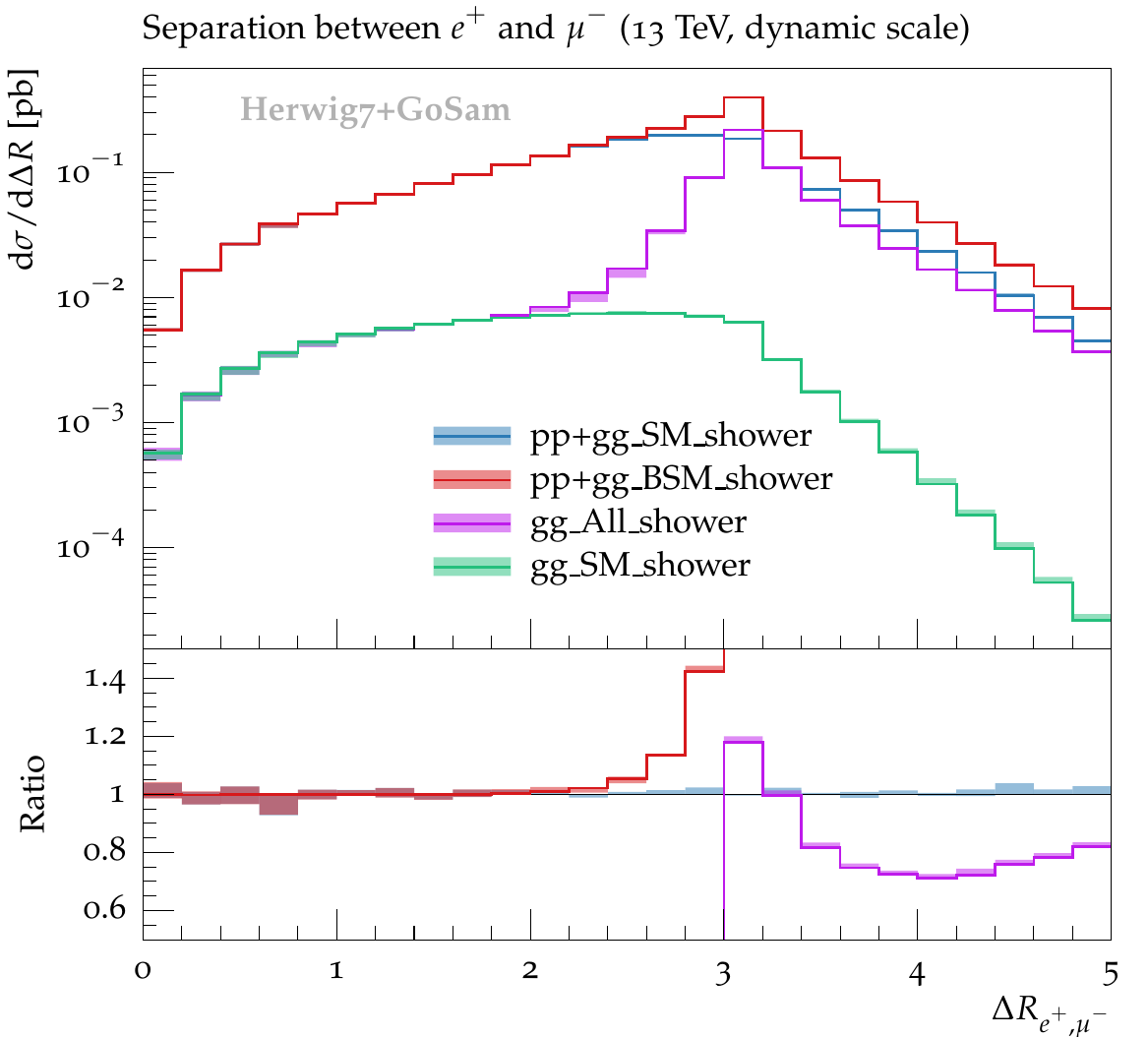}} \hfill
\subfloat[\label{fig:eWmuW_dphi_pp_muQvar}]{ \includegraphics[width=0.45\textwidth]{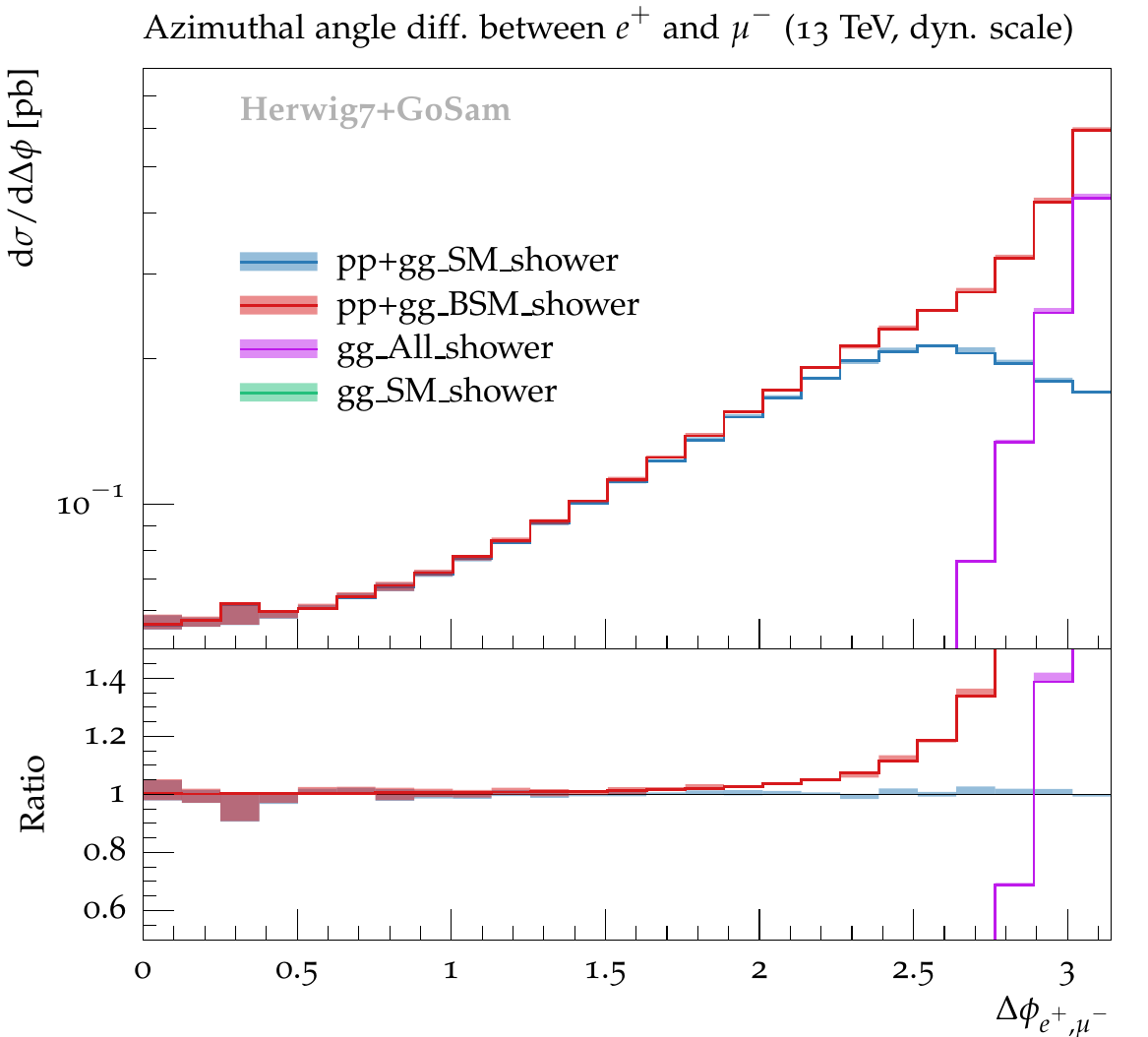}}
\caption{Same as Fig.~\ref{fig:IR_sensitive}, now showing the $\Delta R_{e^+\mu^-}$ and $\Delta\phi_{e^+\mu^-}$ 
distributions.\label{fig:dRdphi_pp}}
\end{figure}

\begin{figure}[htb]
\centering
\subfloat[\label{fig:We_jet1_dR}]{ \includegraphics[width=0.45\textwidth]{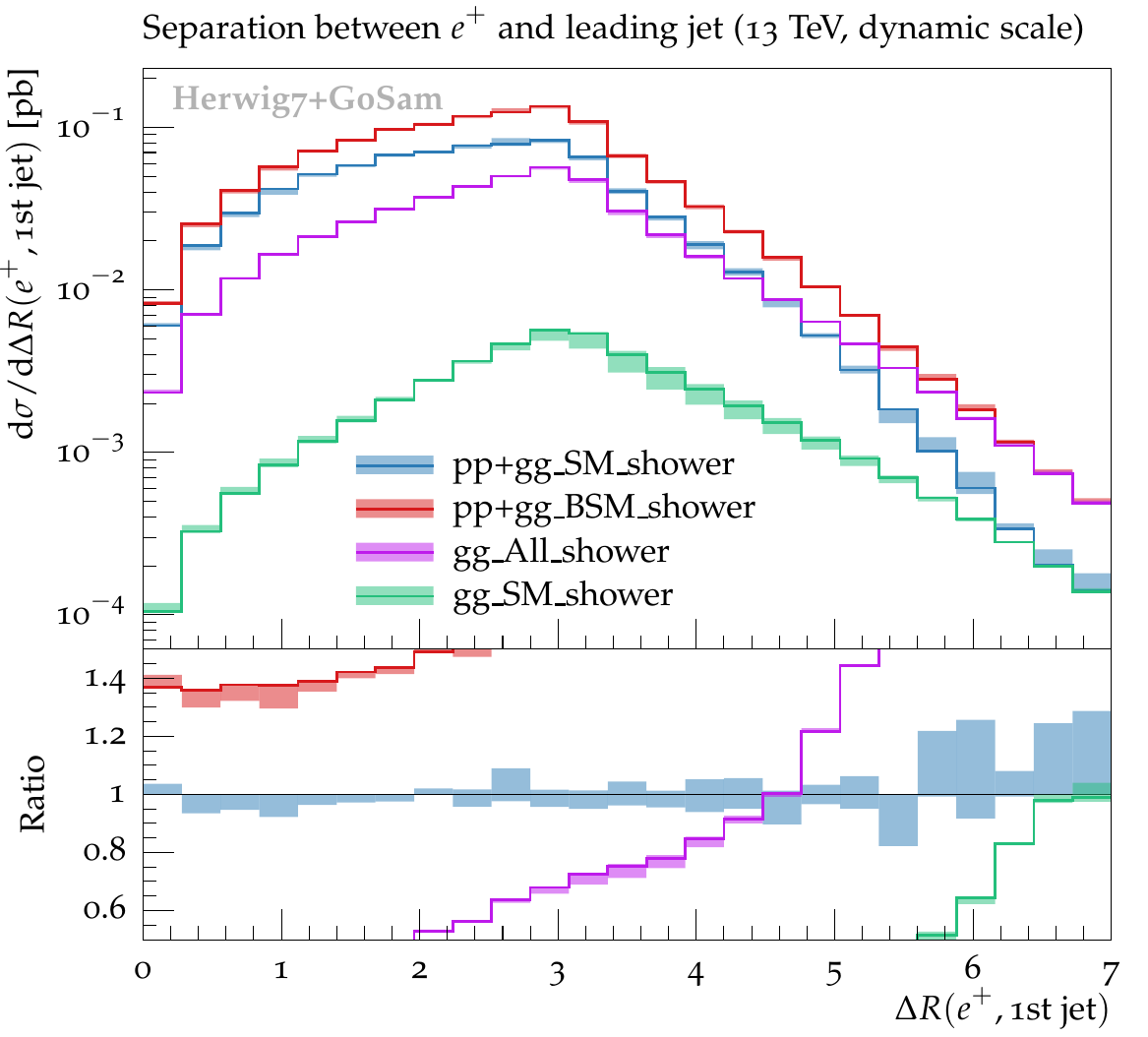}} \hfill
\subfloat[\label{fig:eW_pT_shower}]{ \includegraphics[width=0.45\textwidth]{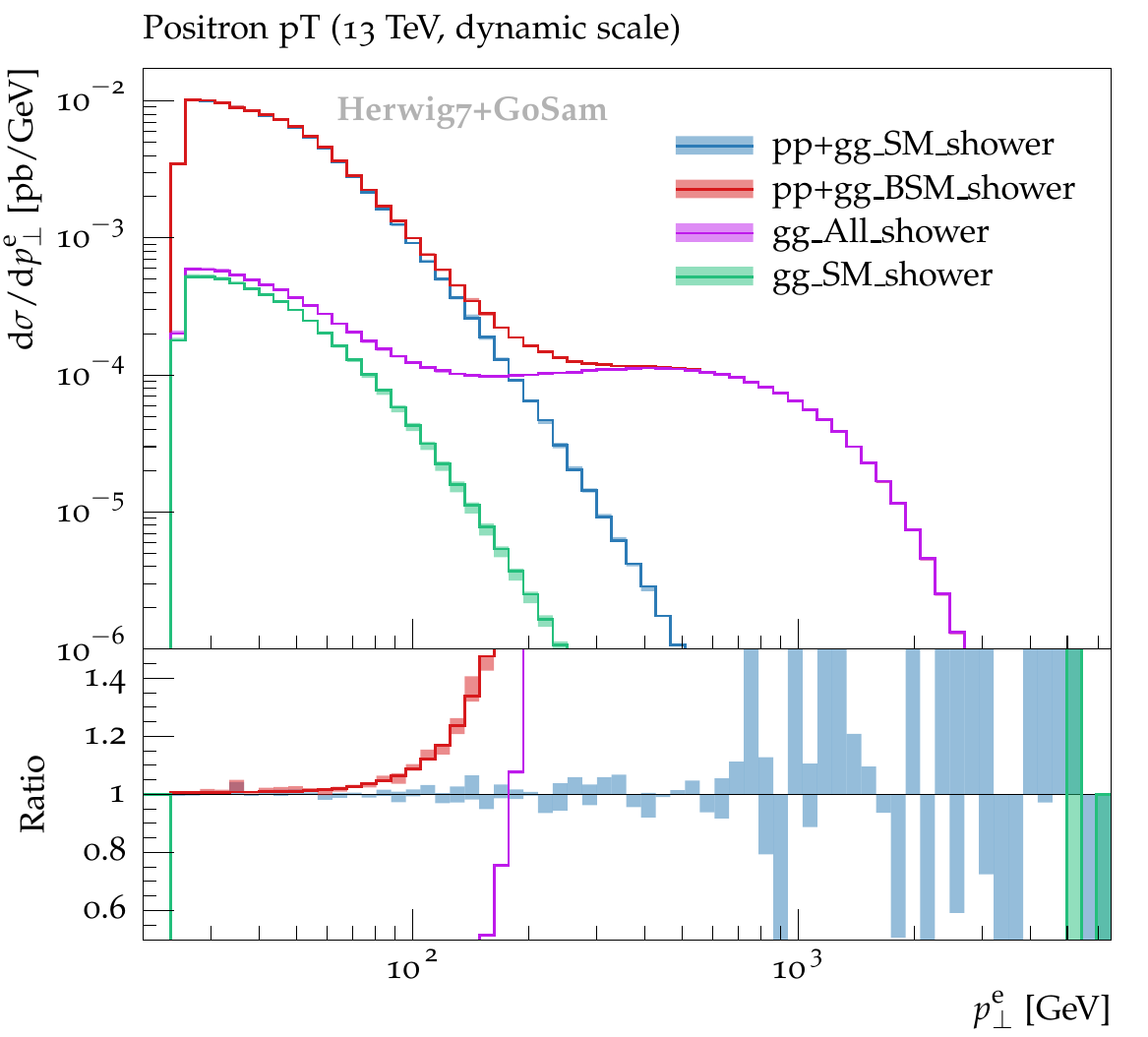}}
\caption{Same as Fig.~\ref{fig:IR_sensitive}, now showing the $\Delta R_{e^+j}$ and $p_{\perp}^{e}$ distributions.
\label{fig:experimentals}}
\end{figure}

A realistic description of observables measured in hadronic collisions
includes effects from a parton shower. We therefore combine our fixed-order NLO
results (supplemented with loop-induced and EFT contributions)
with the \herwig{} angular-ordered parton shower \cite{Gieseke:2003rz}
through the subtractive ({\it i.e.} MC@NLO-like) matching algorithm based on
\cite{Frixione:2002ik,Platzer:2011bc}.\footnote{The loop-induced and EFT
  contributions are treated as LO QCD processes in that regard.}
Uncertainties in the shower are mainly quantified by varying the hard shower
scale $\mu_Q$ which provides a reliable estimate of missing logarithmic orders as well
as the impact of large-angle, hard and thus unreliably modelled emissions. It
also serves as a check to verify the improvements expected from NLO plus
parton-shower matching.

While, as expected, the invariant mass of the reconstructed $W^+W^-$ system is not affected by
the parton shower, a number of other infrared sensitive observables receive
large contributions, both through the NLO real radiation and further subsequent
parton-shower emissions. Typical infrared-sensitive distributions in this case
are the $R$-separation of the two $W$ bosons 
(where in the zero-jet limit $\Delta R$ is  composed solely of a difference in rapidity, while $\Delta \phi=\pi$),
as well as the
transverse momentum of the reconstructed $W^+W^-$ system, shown in
Fig.~\ref{fig:IR_sensitive}. Both observables show the expected behaviour
with respect to additional radiation; in the region $\Delta R < \pi$, both the
NLO real emission as well as shower emissions off the $gg$-induced channel
contribute. The first contribution includes a small shower uncertainty, as
this kinematic range has been improved by the NLO matching. 
Once the BSM contribution to the
$gg$-channel becomes dominant, pure shower
emissions off this sub-process become more important and hence yield a larger
uncertainty.
Ultimately, NLO QCD corrections, or at least a leading-order
multi-jet merging are desirable in this case. Similar features are present in
the transverse momentum of the reconstructed $W^+W^-$ pair. Azimuthal and $R$-separations of the
charged leptons are sensitive to the BSM contribution and very stable against
QCD activity, as shown in Fig.~\ref{fig:dRdphi_pp}.

We finally discuss a few observables which are relevant to the experimental
reconstruction of the $W^+W^-$ final state, particularly lepton-jet separations
and the distribution of missing transverse momentum, displayed in
Fig.~\ref{fig:experimentals}. While shower uncertainties at the level of
10\% are observed, the lepton-jet separation is rather stable against QCD
activity, and BSM contributions only affect the normalization in the
small-$\Delta R$ region; the experimentally required lepton-jet isolation is
thus not introducing any bias. Larger impact is observed on the
transverse momenta of the charged leptons ({\it e.g.} as shown for the positron in Fig.~\ref{fig:experimentals}),
which, however, turn out to be rather stable with respect
to parton-shower scale variations.

\vspace{15pt}
\section{Conclusions and outlook}
\label{sec:conclusion}

The production of electroweak gauge-boson pairs is amongst the most important
signatures at the LHC. These final states are important Higgs-boson decay
channels, and they allow us to study the electroweak sector, with the aim to
reveal the mechanism of electroweak symmetry breaking. 
We have studied the production of a pair of $W$ bosons at NLO QCD, in the light of
additional anomalous couplings. In particular, we have also included the
$gg$-initiated (loop-induced) process $gg\;(\to W^+W^-)\to e^+ \nu_e
\mu^-\bar{\nu}_\mu$, which is formally a
contribution to the NNLO result, but is enhanced due to the large gluon
luminosity at the LHC.  In addition to the Standard-Model $gg$-initiated
contribution, we have included $gg$-initiated contributions stemming from
dimension-eight operators which induce a tree-level coupling between gluons and
electroweak gauge bosons. This possibility has not been discussed in the
literature so far. Their presence leads to an interference between the
Standard-Model $gg$-induced one-loop amplitude and a tree-level amplitude mediated by
dimension-eight operators.

We have discussed their effects on a variety of
important observables.  We have found that the presence of dimension-eight
operators can lead to substantial effects in the high-energy tail of the
distributions, which can be used by the LHC experiments to constrain the
parameter space for the associated effective couplings.

Furthermore, we have
investigated the importance of heavy (SM) quarks in the loop-induced process,
leading to corrections of up to 10\%, depending on cuts and center-of-mass
energy. 

Finally, by combining our fixed-order results with the \herwig{} angular-ordered parton shower,
we have studied the effects of a parton shower, including variations of the hard shower scale,
on the leptonic observables and on observables related to the reconstructed $W^+W^-$ system.

\vspace{15pt}
\section*{Acknowledgements}
NG, GH and JFvSF would like to thank the \gosam{} collaboration for various work on code improvement.
Likewise JB, SG, SP and CR would like to thank the {\sc{Herwig}} collaboration.
This research was supported in part by the Research Executive Agency (REA) of
the European Union under the Grant Agreements PITN-GA2012316704
(HiggsTools) and PITN-GA-2012-315877 (MCnetITN).
SP acknowledges support by a FP7 Marie Curie Intra European
Fellowship under Grant Agreement PIEF-GA-2013-628739.
CR acknowledges partial support by the German Federal Ministry of Education and Research (BMBF).
NG was supported by the Swiss National Science Foundation under contract PZ00P2\_154829.

 \bibliographystyle{JHEP}
 \bibliography{refs_wwjet}

\end{document}